\documentclass[acmtog, timestamp]{acmart}
\usepackage{prettyref}
\usepackage{wrapfig}
\usepackage{multirow}
\usepackage{booktabs}
\usepackage{algorithm}
\usepackage{algpseudocode}
\usepackage{enumitem}
\algnewcommand{\LineComment}[1]{\State \(\triangleright\) #1}
\algdef{SE}[DOWHILE]{Do}{doWhile}{\algorithmicdo}[1]{\algorithmicwhile\ #1}
\newcommand*{\colorboxed}{}
\def\colorboxed#1#{%
  \colorboxedAux{#1}%
}
\newcommand*{\colorboxedAux}[3]{%
  \begingroup
    \colorlet{cb@saved}{.}%
    \color#1{#2}%
    \boxed{%
      \color{cb@saved}%
      #3%
    }%
  \endgroup
}

\newrefformat{Fig}{Figure~\ref{#1}}
\newrefformat{fig}{Figure~\ref{#1}}
\newrefformat{par}{Section~\ref{#1}}
\newrefformat{appen}{Appendix~\ref{#1}}
\newrefformat{sec}{Section~\ref{#1}}
\newrefformat{sub}{Section~\ref{#1}}
\newrefformat{table}{Table~\ref{#1}}
\newrefformat{alg}{Algorithm~\ref{#1}}
\newrefformat{Alg}{Algorithm~\ref{#1}}
\newrefformat{Def}{Definition~\ref{#1}}
\newrefformat{Thm}{Theorem~\ref{#1}}
\newrefformat{step}{Step~\ref{#1}}
\newrefformat{ln}{Line~\ref{#1}}
\newrefformat{eq}{Equation~\ref{#1}}
\newrefformat{pb}{Problem~\ref{#1}}
\newrefformat{it}{Item~\ref{#1}}
\newrefformat{te}{Term~\ref{#1}}
\def\Eqref Eq:#1:{\eqref{eq:#1}}
\newrefformat{Eq}{Equation~\Eqref#1:}

\newcommand{\shortref}[2]{Fig.\ref{#1}#2}

\newcommand{\E}[1]{\mathbf{#1}}
\newcommand{\TE}[1]{\textbf{#1}}

\newcommand{\FPP}[2]{\frac{\partial{#1}}{\partial{#2}}}
\newcommand{\FPPT}[2]{\frac{\partial^2{#1}}{\partial{#2}^2}}

\newcommand{\FPPM}[3]{\frac{\partial^2{#1}}{\partial{#2}\partial{#3}}}

\newcommand{\TWO}[2]{\left(\setlength{\arraycolsep}{1pt}\begin{array}{cc}{#1} & {#2}\end{array}\right)}
\newcommand{\TWOC}[2]{\left(\setlength{\arraycolsep}{1pt}\begin{array}{c}#1 \\ #2\end{array}\right)}

\newcommand{\THREE}[3]{\left(\setlength{\arraycolsep}{1pt}\begin{array}{ccc}{#1} & {#2} & {#3}\end{array}\right)}

\newcommand{\THREEC}[3]{\left(\setlength{\arraycolsep}{1pt}\begin{array}{c}#1 \\ #2 \\ #3\end{array}\right)}

\newcommand{\FOUR}[4]{\left(\setlength{\arraycolsep}{1pt}\begin{array}{cccc}{#1} & {#2} & {#3} & {#4}\end{array}\right)}
\newcommand{\FOURC}[4]{\left(\setlength{\arraycolsep}{1pt}\begin{array}{c}#1 \\ #2 \\ #3 \\ #4\end{array}\right)}

\newcommand{\MTT}[4]{\left(\setlength{\arraycolsep}{1pt}\begin{array}{cc}#1 & #2 \\ #3 & #4\end{array}\right)}

\newcommand{\MDD}[3]{\left(\setlength{\arraycolsep}{1pt}\begin{array}{ccc}#1 & & \\ & #2 & \\ & & #3\end{array}\right)}

\newcommand{\fmin}[1]{\underset{#1}{\E{min}}}
\newcommand{\fmax}[1]{\underset{#1}{\E{max}}}
\newcommand{\argmin}[1]{\underset{#1}{\E{argmin}}}

\newcommand{\traj}{\mathcal{Q}_K}
\newcommand{\DMPW}[1]{\mathcal{W}_{#1}}
\newcommand{\fext}[2]{{\mathcal{E}_{#1}^{#2}}}
\renewcommand{\fint}[2]{{\mathcal{F}_{#1}^{#2}}}

\newcommand{\ENG}[1]{E_{\text{#1}}}
\newcommand{\ENGS}[2]{E_{\text{#1}}^{\text{#2}}}
\newcommand{\COEF}[1]{C_{\text{#1}}}
\newcommand{\COEFS}[2]{C_{\text{#1}}^{\text{#2}}}
\newcommand{\CROSS}[1]{[#1]}
\newcommand{\MASS}[1]{M_\text{#1}}

\newcommand{\refSuppDeriv}{\prettyref{appen:VAID} }
\newcommand{\refSuppKinetic}{\prettyref{appen:KCA} }
\newcommand{\refSuppOptimizer}{\prettyref{appen:OA} }
\newcommand{\refSuppPhysVio}{\prettyref{appen:PV} }

\usepackage{tikz}
\usetikzlibrary[arrows,backgrounds,decorations.pathmorphing,positioning,fit,petri]

\newcommand{\changed}[1]{\textcolor{black}{#1}}
\newenvironment{changedBlk}{\color{black}}{}

\setcitestyle{square}

\setcopyright{none}

\begin{document}
\title{\changed{Active Animations of Reduced Deformable Models with Environment Interactions}} 
\author{Zherong Pan}
\affiliation{%
  \institution{University of North Carolina at Chapel Hill}
  \streetaddress{SN334, Columbia Street}
  \city{Chapel Hill}
  \state{NC}
  \postcode{27514}
  \country{USA}}
\author{Dinesh Manocha}
\affiliation{%
  \institution{University of North Carolina at Chapel Hill}
  \streetaddress{SN334, Columbia Street}
  \city{Chapel Hill}
  \state{NC}
  \postcode{27514}
  \country{USA}}
\renewcommand\shortauthors{Pan, et al}

\begin{abstract}
We present an efficient spacetime optimization method to automatically generate animations for a general volumetric, elastically deformable body. Our approach can model the interactions between the body and the environment and automatically generate active animations. We model the frictional contact forces using contact invariant optimization and the fluid drag forces using a simplified model. To handle complex objects, we use a reduced deformable model and present a novel hybrid optimizer to search for the local minima efficiently. This allows us to use long-horizon motion planning to automatically generate animations such as walking, jumping, swimming, and rolling. We evaluate the approach on different shapes and animations, including deformable body navigation and combining with an open-loop controller for realtime forward simulation.
\end{abstract}

%
%

%
%

\begin{teaserfigure}
\scalebox{1.03}{
\includegraphics[width=0.96\textwidth]{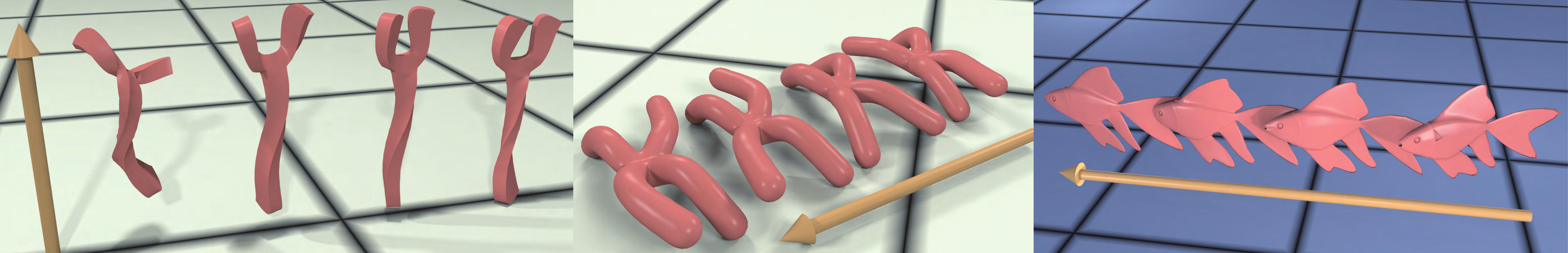}
\put(-330,5){(a)}
\put(-185,5){(b)}
\put(- 20,5){(c)}}
\caption{\label{fig:Teaser} Active deformable animations automatically generated by our approach: A letter T jumping (a), a spider walking (b), and a fish swimming (c). The reduced configuration spaces of these deformable bodies have $5-15$ DOFs.  We present an efficient spacetime optimization formulation that takes into account physics constraints and environmental interactions.}
\end{teaserfigure}
\maketitle

\section{Introduction}\label{sec:intro}
Physically-based deformable animation is a well-studied problem in computer graphics and related areas. Early methods such as \cite{Terzopoulos:1987:EDM:37401.37427,Muller:2004:IVM:1006058.1006087} focus on passive animations using numerical simulations. These techniques are widely used to generate plausible simulations of clothes~\cite{Bridson:2002:RTC:566654.566623}, plants~\cite{Barbic:2011:RLS:1964921.1964986}, human tissues~\cite{Chentanez:2009:ISS:1576246.1531394}, etc. Such passive animations are frequently used in movies and games to increase the realism. On the other hand, generating controlled or active deformable body animations~\cite{Tan:2012:SBL:2185520.2185522,Coros:2012:DOA:2185520.2185565,Kim:2011:FSS:2019627.2019640} is considered more challenging, especially when a deformable body's movements are governed by physics-based constraints. In such cases, additional control inputs, such as keyframes or rest shapes, need to be determined based on a deformable body's interactions with the environment in order to generate the animation. This can be computationally challenging for deformable bodies with a high number of degrees of freedom (DOFs). To simplify the problem, previous methods \cite{Kim:2011:PCS:2019406.2019415,Hahn:2012:RP:2185520.2185568,Harmon:2013:SIL:2461912.2461922,Liu:2013:SCS:2508363.2508427,Xu:2016:PSD:2897824.2925916} partition the deformable body's DOFs into controlled DOFs and uncontrolled DOFs. In practice, prior techniques specify the trajectories of controlled DOFs using manual keyframes and use physics-based simulation algorithms to generate movements corresponding to uncontrolled DOFs, i.e., the secondary dynamics. Such techniques are widely used for physical character rigging. In general, it is hard to generate controlled deformable body animations without user intervention or specifications. The animators not only need to manually partition the DOFs into controlled DOFs and the uncontrolled DOFs, but they also need to specify the movements of the controlled DOFs.

\begin{changedBlk}
\TE{Main Results}: We present a new method for active deformable body animations. The input to our method is a volumetric mesh representation of the body, a specification of the environment, and a high-level objective function that is used to govern the object's movement. Our algorithm can automatically compute active animations of the deformable body and can generate motions corresponding to walking, jumping, swimming, or rolling, as shown in \prettyref{fig:Teaser}. We compute the animations using a novel spacetime optimization algorithm and formulate the objective function taking into account dynamics constraints as well as various interactions with the environment. These include collisions, frictional contact forces and fluid drag forces. Compared with keyframe-based methods, we use objective functions to control the animation. In practice, these objective functions are more general and easier for the user to specify. For example, to generate walking animation, user can just specify the target walking speed, instead of manually specifying the walking poses corresponding to different timesteps. Furthermore, our approach can be easily combined with partial keyframe data to provide more user control.

 Some of the novel components of our work include: 
\begin{itemize}
\item A spacetime optimization formulation using reduced deformable models (\prettyref{sec:objective}), which takes into account environment interactions.
\item A hybrid spacetime optimization algorithm (\prettyref{sec:optimization}), which is more than an order of magnitude faster than previous spacetime optimization methods. 
\item We combine our spacetime optimization algorithm with dynamic movement primitives (DMP), which have been used in robotics~\cite{schaal2006dynamic}. DMP improves the performance of our algorithm in terms of avoiding suboptimal solutions. Furthermore, we present a two-stage animation framework. During the first stage, we compute the animation trajectories using DMP as a prior. These animations are then tracked and composed together at realtime using DMP as a controller (\prettyref{sec:results}).
\end{itemize}

We demonstrate the benefits of our method by evaluating its performance on different complex deformable bodies with thousands of vertices and $5-15$ DOFs in different environments (\prettyref{sec:results}). For underwater swimming, we use DMP as an open-loop controller to generate realtime swimming animations (\prettyref{fig:swim}). For contact-rich locomotion, the optimized animations are tracked at realtime using a feedback controller (\prettyref{fig:trackT}). Finally, we formulate keyframe-based control as a special case of our method and show animations controlled using partial keyframes and high-level control objectives (\prettyref{fig:dinoWalk}).
\end{changedBlk}
\section{Related Work}\label{sec:related}
Our work is inspired by prior work on passive/active deformable body animations and model reduction. In this section, we give a brief overview of related work.
\begin{figure}[t]
\centering
\includegraphics[width=0.48\textwidth]{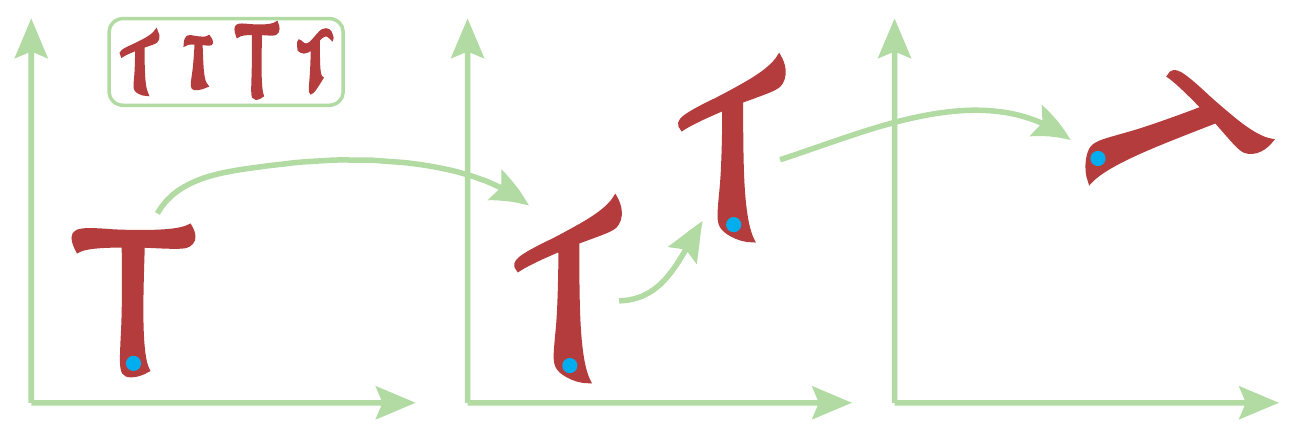}
\put(-210,10){$v^j(0)$}
\put(-180,40){$u$}
\put(-115,20){$c$}
\put(-70,50){$w$}
\put(-40,37){$v^j(q)$}
\put(-230,45){(a)}
\put(-175,70){(b)}
\put(-125,70){(c)}
\put(-45,70){(d)}
\caption{\label{fig:reduced} The deformable body is represented as a triangle mesh (a). With a deformation bases set (b), its position is parameterized by local deformation $u$, a global rigid translation $c$ (c), and rotation $w$ (d). The Euclidean coordinates of a vertex $v^j$ (blue dot) can be recovered by transformation function $v^j(q)$.}
\end{figure}
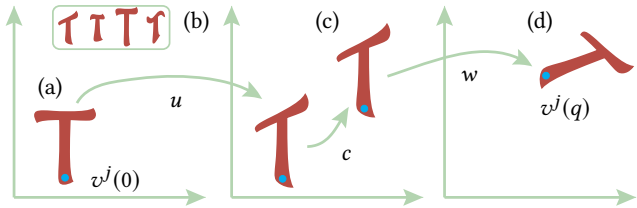

\TE{Passive Deformable Body Animation} has been an active area of research for more than three decades. The most popular deformable model, especially for deformable bodies without skeletons, is the finite element method (FEM) \cite{Terzopoulos:1987:EDM:37401.37427,Irving:2006:THI:1140961.1140963}. These methods of deformable body modeling have computational complexity that is superlinear in the number of discrete elements, and they therefore are not suitable for interactive applications. Many deformable bodies such as human bodies and animals have embedded rigid skeletons. Robust methods such as \cite{Capell:2002:ISD:566570.566622,Kim:2011:FSS:2019627.2019640,Hahn:2012:RP:2185520.2185568} have been proposed to model skeletons' interactions with soft tissues. In this paper, we use FEM to model a deformable body.

\TE{Active Deformable Body Animation} is used by animators or artists to direct the animation while satisfying the physics-based constraints. Early works in this area \cite{Bergou:2007:TTD:1275808.1276439,Barbic:2009:DOA:1576246.1531359,Barbic:2012:IED:2185520.2185566,Schulz:2014:ADO:2601097.2601156} try to make the deformable body follow a user-provided animation by applying external forces. However, deformable bodies in real life, such as worms, snakes, and fishes, can only move themselves by generating internal forces. To respect this property, \cite{Kim:2011:FSS:2019627.2019640,Tan:2012:SBL:2185520.2185522,Coros:2012:DOA:2185520.2185565} control virtual deformable bodies to follow a given animation by applying internal forces only. Our work can be considered as complimentary to these methods. We generate animations that can be used as input to these previous methods, with a focus on reduced deformable models. \changed{Deformable body control methods can also be categorized based on the underlying user interfaces:  \cite{Bergou:2007:TTD:1275808.1276439,Barbic:2009:DOA:1576246.1531359,Barbic:2012:IED:2185520.2185566,Schulz:2014:ADO:2601097.2601156,Kim:2011:FSS:2019627.2019640} require the user to specify  a set of spacetime keyframes, while \cite{Tan:2012:SBL:2185520.2185522,Coros:2012:DOA:2185520.2185565} and our approach specify the goals for controlling the animation using objective functions.}

\changed{\TE{Spacetime Optimization} Many techniques for deformable body animations are based on spacetime optimization \cite{Witkin:1988:SC:54852.378507}. Solving these optimization problems can be challenging due to the high-dimensional search space. Some algorithms parameterize the search space using low-dimensional representations such as splines \cite{Hildebrandt:2012:ISC:2185520.2185567} and functional spaces \cite{7487091}. Another challenging issue is handling of non-smoothness constraints in the optimization formulation, due to environment interactions corresponding to collisions and contacts. Previous work either use sampling-based methods \cite{Xu:2016:PSD:2897824.2925916}, complementarity constrained optimizations \cite{Peng:2017:DDL:3072959.3073602}, or a smooth variant of the contact model \cite{Mordatch:2012:DCB:2185520.2185539,Mordatch:2013:AHL:2508363.2508365}. In our work, we handle the  high-dimensionality using reduced deformable models and solve the optimization problem using a hybrid method.}

\TE{Model Reduction} is a practical method for fast deformable body simulations. It is based on the observation that only visually salient deformations need to be modeled. The earliest reduced model is based on Linear Modal Analysis (LMA) \cite{Pentland:1989:GVM:74334.74355,Hauser:2003:IDU}, which is only accurate for infinitesimal deformations. Methods for non-linear and large deformations have been proposed in \cite{1359737,Barbic:2005:RSI:1186822.1073300,An:2008:OCE:1457515.1409118}. In this paper, we use the rotation-strain space dynamic model \cite{Pan:2015:SDS:2816795.2818090} because it can preserve the key characteristics of deformable bodies with a lower-dimensional configuration space representation. However, our method can also be used with other reduced dynamic models.
\section{Problem Formulation}\label{sec:problem}
In this section, we formulate the problem of generating deformable body animation as a spacetime optimization. Our method searches in the space of deformable body animations with a fixed number of $K$ timesteps. We denote an animation trajectory as: $\traj=\THREE{q_1}{\cdots}{q_K}$, where each state vector $q_i$ uniquely determines the position of a deformable body at time instance $i\Delta t$, where $\Delta t$ is a fixed timestep size.

\begin{changedBlk}
\subsection{\label{sec:IO} Input $\&$ Output}
In \prettyref{sec:optimization}, we present an efficient optimizer to robustly search for the animation trajectory $\traj$. Our overall algorithm takes the following components as an input:
\begin{itemize}
    \item A volumetric mesh representation of the deformable body with $V$ vertices $\THREE{v^1}{\cdots}{v^V}$.
    \item A specification of the environment, including type of the environment (in water, or on the ground) and parameters of the environment (e.g., drag force coefficient in water, or contact friction coefficient on the ground).
    \item The form of high level objective $\ENG{obj}$ and its parameters. For example, in order for a deformable body to walk to a target position, $\ENG{obj}$ will penalize the distance between current center of mass $c$ and the target position, and its parameters correspond to the target position's coordinates.
\end{itemize}
\end{changedBlk}

\subsection{Objective Function for Spacetime Optimization}
By spacetime optimization, we assume that the desired deformation body animation corresponds to the local minima of an objective function. As a result, this objective function must encode all the requirements for a physically correct and plausible deformable body animation. We model these requirements by taking four different energy terms into account in $E(\traj)$:
\begin{equation}
\label{eq:obj}
E(\traj)=\ENG{phys}+\ENG{obj}+\ENG{env}+\ENG{hint}.
\end{equation}
\begin{figure}[t]
\centering
\includegraphics[width=0.48\textwidth]{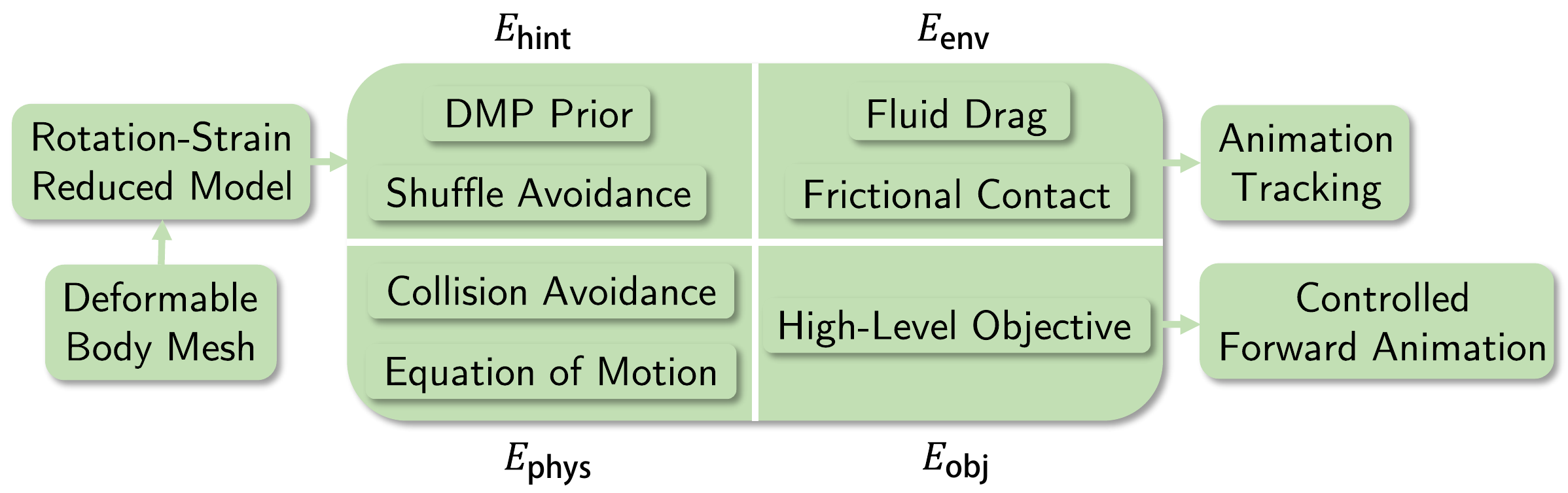}
\caption{\label{fig:flowchart} We highlight various components used in our spacetime optimization algorithm that can also be used for animation tracking and controlled forward animation.}
\end{figure}
As outlined in \prettyref{fig:flowchart}, \changed{the first term $\ENG{phys}$ models all the shape changes that occur within the deformable body, i.e., the dynamics that result from the internal forces. It penalizes any violation in the deformable body's equation of motion (\prettyref{sec:physics}) and any collisions between different parts of the deformable body (\prettyref{sec:coll}). The second term $\ENG{obj}$ is a task-dependent objective function specified by user (see \prettyref{sec:IO}). The environmental force term $\ENG{env}$ (\prettyref{sec:environment}) models all the dynamic interactions between the deformable object and the environment, i.e. due to the external forces. It also penalizes any violation in the constraints that the environmental forces, such as frictional contact forces, must satisfy.} Finally, the last term $\ENG{hint}$ (\prettyref{sec:hint}) guides the optimizer to avoid local minima that may result in less plausible animations.

\subsection{Configuration Space Parametrization}
Although our method can work with any parametrization of the deformation body's configuration $q$, different parametrizations result in drastically different computational cost. A straightforward method is to use volumetric meshes with $V$ vertices $\THREE{v^1}{\cdots}{v^V}$ and define $q$ as all the vertices' Euclidean coordinates. In this case, the dimension of the configuration space, $|q|$, scales linearly with the number of vertices and may be several thousands for moderately complex deformable models. Using optimization algorithms in such a large search space is only practical for very short animations. Indeed, \cite{Tan:2012:SBL:2185520.2185522,Bergou:2007:TTD:1275808.1276439} used this vertex-based parametrization for tracking deformable body animation in a frame-by-frame manner, i.e., $K=1$.

Instead, we represent the configuration of a deformable body using a rigid-coupled reduced model defined as:
\begin{equation}
\label{eq:RS}
q=\THREE{u}{c}{w}^T,
\end{equation}
where $u$ parametrizes the deformable body's non-rigid deformations in its local frame of reference. This is complemented with a rigid transformation in the world coordinates parametrized using a global translation $c$ and rotation $w$, as illustrated in \prettyref{fig:reduced}. By using a precomputed dataset of deformation bases, the dimension of local deformation $|u|$ in \prettyref{eq:RS} is usually no more than $20$. Moreover, methods such as cubature approximation \cite{An:2008:OCE:1457515.1409118} and fast sandwich transform (FST) \cite{Kim:2011:PCS:2019406.2019415} can be used to efficiently recover a vertex $v^j$'s Euclidean coordinates using the transformation function $v^j(q)$. This transformation function can take a different form depending on the underlying reduced dynamic models. \changed{We refer readers to \prettyref{sec:analyze} for more analysis in terms of combining our method with different reduced dynamic models.} A widely-known model is the reduced StVK \cite{Barbic:2005:RSI:1186822.1073300}. Instead, we use the recently proposed rotation-strain (RS) space dynamic model \cite{Pan:2015:SDS:2816795.2818090} because it achieves comparable results with a lower-dimensional configuration space, i.e., a smaller $|u|$. We provide details about the computation of $v(q)$ in \refSuppDeriv. We denote the reconstructed Euclidean coordinates representation as:
\begin{eqnarray*}
\bar{q}(q)=\FOUR{v^1(q)}{v^2(q)}{\cdots}{v^V(q)}^T.
\end{eqnarray*}
These formulations make it computationally tractable to numerically optimize a complex nonlinear function $E(\traj)$. Moreover, \prettyref{eq:RS} is very convenient in terms of formulating our objective functions $\ENG{obj}$. For example, we could use a function in $c$ to direct a deformable body to walk to a specific position, or a function in $w$ to specify that a deformable body should stay balanced.
\section{Objective Terms}\label{sec:objective}
In this section, we present details of the objective function used for spacetime optimization.

\subsection{Physics-Based Constraints}
The first term $\ENG{phys}$ penalizes any violation of the equations of motion (EOM), and our formulation is similar to prior work \cite{Barbic:2009:DOA:1576246.1531359,Barbic:2012:IED:2185520.2185566}. In addition, we also penalize any self-penetrations or collisions with static obstacles. Altogether, $\ENG{phys}$ is represented as:
\begin{eqnarray*}
\ENG{phys}(\traj)=\sum_{i=2}^{K-1}\ENG{eom}(q_{i-1},q_i,q_{i+1})+\ENG{coll}(q_i)+\ENG{self}(q_i).
\end{eqnarray*}

\subsubsection{Equations of Motion}\label{sec:physics}
Since $q_i$ only represents a deformable body's position, $\ENG{eom}$ models the dynamic behavior using 3 consecutive frames. one advantage of this formulation is that we can use a position-based large timestep integrator \cite{Hahn:2012:RP:2185520.2185568} to formulate our EOM. 
An implicit-Euler scheme determines $q_{i+1}$ from $q_{i-1},q_i$ using the following optimization formulation:
\begin{eqnarray*}
q_{i+1}&=&\argmin{q}\left[\frac{A(q)^TMA(q)}{2\Delta t^2}+P(q)-\fint{i}{T}u-\fext{i}{T}\bar{q}(q)\right] \nonumber \\
A(q)&\triangleq&\bar{q}(q)-2\bar{q}(q_{i})+\bar{q}(q_{i-1}) \nonumber,
\end{eqnarray*}
where $M$ is the mass matrix constructed from the volumetric mesh of the deformable body using FEM, $\fint{}{}$ represents the internal control forces and $\fext{}{}$ corresponds to the environmental forces such as gravitational forces, fluid drag forces, and frictional contact forces. $P$ is the elastic potential energy, and we model this energy term using the rotation-strain space linear elastic energy $P(q)=u^T\mathcal{K}u/2$, where $\mathcal{K}$ is the isotropic stiffness matrix. Even with an arbitrarily large $\Delta t$, the above time integrator is always stable. The term $\ENG{eom}$ is simply defined as the norm of gradient:
\begin{eqnarray}
\label{eq:eom}
&&\ENG{eom}(q_{i-1},q_i,q_{i+1})  \\
&=&\frac{1}{2}\|\FPP{\bar{q}}{q}^T(q_{i+1})MA(q_{i+1})+\FPP{\left[P-\fint{i}{}^Tu-\fext{i}{}^T\bar{q}\right]}{q}(q_{i+1})\|^2 \nonumber.
\end{eqnarray}

\subsubsection{Collision Avoidance}\label{sec:coll}
Collision handling is regarded as a challenging problem in terms of deformable body simulation. In our method, we use two terms to approximately avoid collisions. For collisions with static obstacles, we formulate an energy term as:
\begin{eqnarray*}
\ENG{coll}(q)=\frac{\COEF{coll}}{2}\sum_{j=1}^{V}\E{max}(\E{dist}(v^j(q)),0)^2,
\end{eqnarray*}
where $\E{dist}(v)$ is the signed distance from a vertex's position to static obstacles. To evaluate this function for vertices, we precompute a signed distance field for the static obstacles.

\setlength{\columnsep}{4pt}
\begin{wrapfigure}{r}{4.5cm}
\centering
\scalebox{0.9}{
\includegraphics[width=0.25\textwidth]{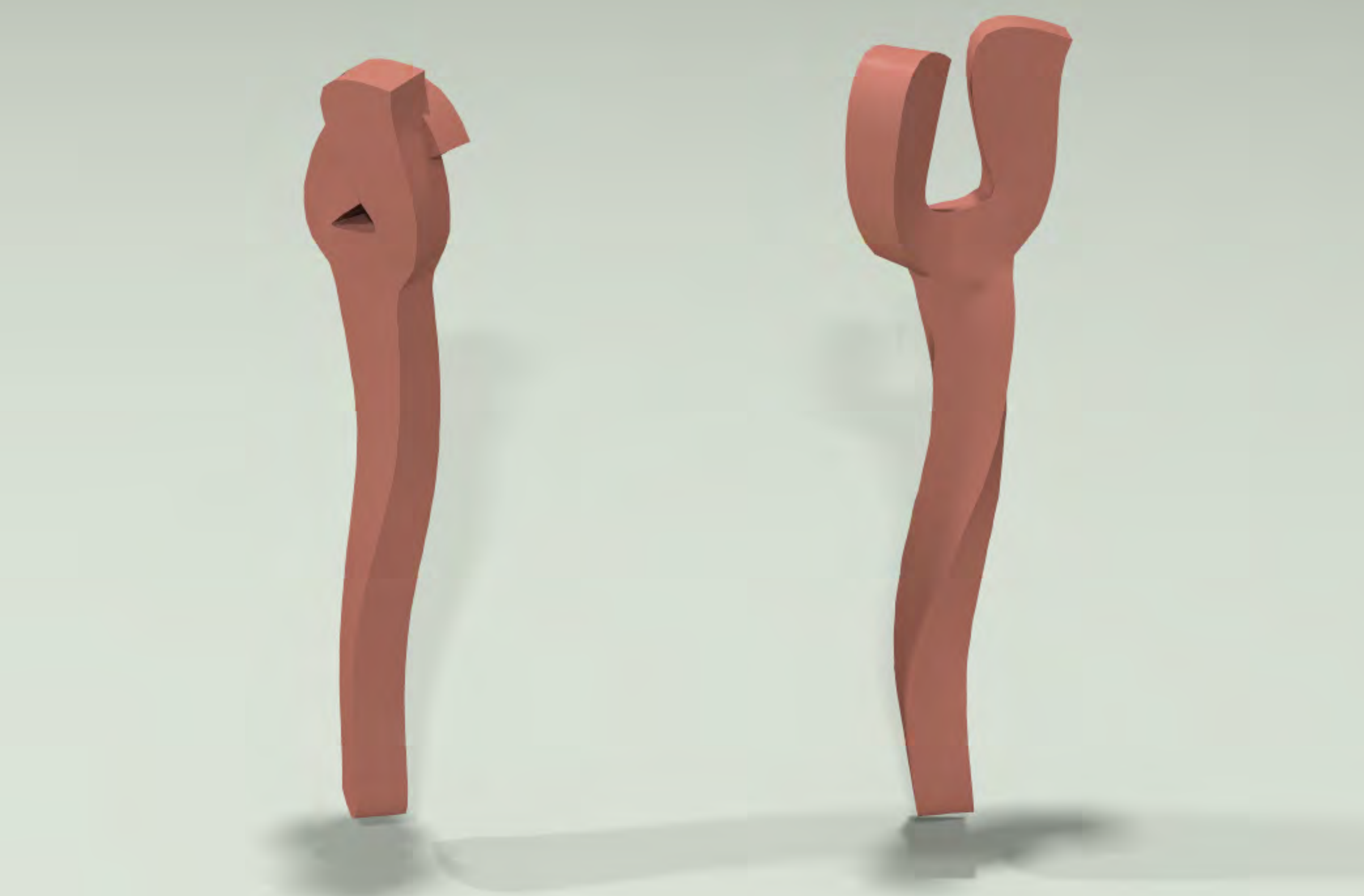}
\put(-80,70){(a)}
\put(-20,70){(b)}}
\caption{\label{fig:CCD} To resolve collisions between thin components (a), we use approximate continuous collision handling (b).}
\end{wrapfigure}
Handling self-collisions is even more challenging. In order to generate animations such as walking and jumping, many deformable bodies have thin structures that function as legs. Successful handling of self-collisions between such thin structures usually requires continuous collision detection (CCD), as illustrated in \prettyref{fig:CCD}. We make use of our reduced representation and use an approximate CCD scheme. Given a configuration $q=\THREE{u}{c}{w}$ that has self-penetrations, we first search for colliding pairs of vertices by reconstructing $\bar{q}(q_i)$ from $q_i$ and run a conventional discrete collision detection. As shown in \cite{Barbic:2010:SSC:1778765.1778818}, considering only vertex-vertex collisions is enough for plausible handling of self-penetrations in reduced-model deformable body animations. Moreover, we observe that self-penetrations are invariant to the global rigid transformation $\TWO{c}{w}$, so we only look at the local deformation component $u$ of $q$. Since we already know that the undeformed configuration, i.e., $u=0$, has no self-collisions, we can use a line-search algorithm in $u$ to find the largest $\eta\in(0,1]$ such that $\eta u$ has no self-collisions. For each pair of vertices, $v^a(u)$ and $v^b(u)$, in collision, we add an energy term:
\begin{eqnarray*}
\ENG{self}^{ab}(u)&=&\E{min}((v^a(u)-v^b(u))^Td^{ab},0)^2  \\
d^{ab}&\triangleq&\frac{v^a(\eta u)-v^b(\eta u)}
{\|v^a(\eta u)-v^b(\eta u)\|},
\end{eqnarray*}
where we use $d^{ab}$ as an approximate direction of separation. $\ENG{self}$ is then defined as:
\begin{eqnarray*}
\ENG{self}(u)&=&\frac{\ENG{self}}{2}\sum_{a=1}^{V}\sum_{b=1}^{V}\ENG{self}^{ab}(u)I^{ab},
\end{eqnarray*}
where the last $I^{ab}$ is an indicator of whether $v^a(q)$ and $v^b(q)$ are in collision.
\subsection{Environmental Force Model}\label{sec:environment}
Since we allow only internal forces $\fint{}{}$ as the control input, a deformable body must make use of external environmental forces $\fext{}{}$ to move around. We consider two kinds of environmental forces: frictional contact forces and fluid drag forces. The frictional contact forces are used for generating contact-rich animations such as walking, balancing, rolling or jumping. The fluid drag forces are used for underwater swimming.

\subsubsection{Frictional Contact Force Model}
To model the frictional contact forces, we use contact invariant optimization (CIO) \cite{Mordatch:2012:DCB:2185520.2185539,Mordatch:2013:AHL:2508363.2508365} and leave external forces $\fext{}{}$ as an additional optimizable variable. However, $\fext{}{}$ must satisfy two additional constraints. First, the contact force on vertex $v^j$, $\fext{}{j}$ should lie inside the frictional cone, we have:
\begin{eqnarray}
\label{eq:cone}
\|\fext{i}{j}_\parallel\| \leq \mu \fext{i}{j}_\perp,
\end{eqnarray}
where $\parallel$ and $\perp$ are the tangent and normal component of the contact force, respectively, and $\mu$ is the frictional coefficient. The big advantage of CIO is that it allows the optimizer to jointly search for both contact forces and contact points by introducing the so-called contact-integrity term defined as:
\begin{eqnarray}
\label{eq:CIO}
&&\ENG{env}(q_i,q_{i-1})=\ENG{env}\sum_{i=2}^K\sum_{j=1}^V   \\
&&(\|\E{dist}(v^j(q_i))\|^2+\|(v^j(q_i)-v^j(q_{i-1}))_\parallel\|^2)\|\fext{i}{j}\|^2 \nonumber.
\end{eqnarray}
This term essentially encourages every external force $\fext{}{}$ to have maximal velocity dissipation and every contact point to stay on the contact manifold. Instead, we use a slightly different formulation from \cite{Mordatch:2013:AHL:2508363.2508365} and use a quadratic penalty for $\fext{}{}$. \changed{In this way, the objective function $E(\traj)$ becomes a quadratic function when we are optimizing only with respect to $\fext{}{}$. Together with \prettyref{eq:cone}, we can find the optimal $\fext{}{}$, given $\traj$, by solving a quadratic constrained QP (QCQP) problem. In \prettyref{eq:CIO}, the function $\E{dist}(\bullet)$ returns the closest distance to the environmental obstacles. We compute this efficiently by precomputing a distance field for the environment and we then use a smoothing algorithm \cite{CGF:CGF2058} so that $\E{dist}(\bullet)$ is $C^1$-continuous.}

\subsubsection{Fluid Drag Force Model}
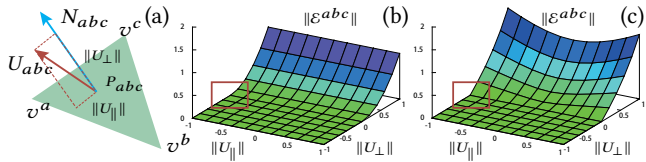
\begin{figure}[t]
\centering
\includegraphics[width=0.48\textwidth]{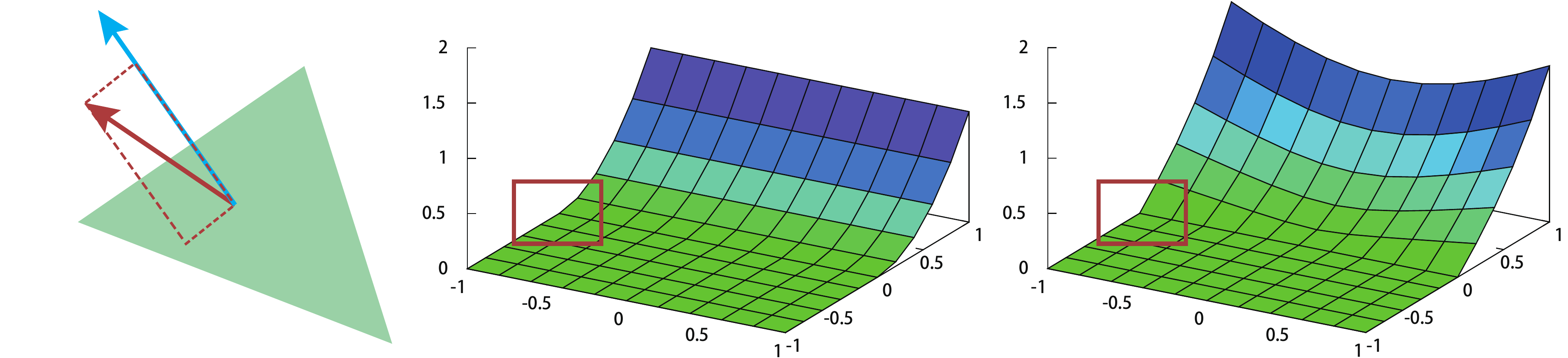}
\put(-242,32){\small$U_{abc}$}
\put(-210,16){\tiny$\|U_\parallel\|$}
\put(-205,26){\tiny$P_{abc}$}
\put(-213,36){\tiny$\|U_\perp\|$}
\put(-222,50){\small$N_{abc}$}
\put(-235,14){\small$v^a$}
\put(-200,45){\small$v^c$}
\put(-182,01){\small$v^b$}
\put(-130,50){\tiny$\|\fext{}{abc}\|$}
\put(-165,1){\tiny$\|U_\parallel\|$}
\put(-110,1){\tiny$\|U_\perp\|$}
\put(-40,50){\tiny$\|\fext{}{abc}\|$}
\put(-75,1){\tiny$\|U_\parallel\|$}
\put(-20,1){\tiny$\|U_\perp\|$}
\put(-190,50){(a)}
\put(-100,50){(b)}
\put(-10,50){(c)}
\caption{\label{fig:drag} Fluid drag force is applied on each surface patch $(v^a,v^b,v^c)$. The force strength depends on the surface normal $N$ and relative velocity $U$ (a). We also plot the force strength with respect to tangential relative velocity $U_\parallel$ and normal relative velocity $U_\perp$. Our new formulation is $C^1$-continuous (b), while the original formulation has a discontinuous gradient (c), especially when the relative velocity is almost tangential (shown with a red rectangle).}
\end{figure}

The fluid drag forces, $\fext{i}{}$, are not free variables but functions of $q_i,q_{i-1}$. \cite{Yuksel:2007:WP:1275808.1276501} used a quadratic drag force model, which is defined as a summation of forces on each triangular surface patch $(v^a,v^b,v^c)$:
\begin{eqnarray}
\label{eq:fluid0}
&&\fext{i}{T}\bar{q}=\fext{}{T}(q_{i+1},q_i)\bar{q}(q_{i+1}) \\
&\triangleq&\COEF{drag}\sum_{(v^a,v^b,v^c)}\E{max}(N_{abc}^TU_{abc},0)U_{abc}^TP_{abc}(q_{i+1}) \nonumber 
\end{eqnarray}
where we have also approximated the surface patch force as a point force on the barycenter $P_{abc}$. Here $N_{abc}$ is the area-weighted normal and $U_{abc}$ is the barycenter's relative velocity against fluid, as illustrated in \prettyref{fig:drag}. Note that \prettyref{eq:fluid0} only takes effect when a surface patch is moving towards the fluid body. However, \prettyref{eq:fluid0} cannot be used by a gradient-based numerical optimizer because the gradient is discontinuous. We propose a continuous model by a slight modification: 
\begin{eqnarray*}
&&\fext{}{T}(q_{i+1},q_i)\bar{q}(q_{i+1})    \\
&\triangleq&\COEF{drag}\sum_{(v^a,v^b,v^c)}\E{max}(N_{abc}^TU_{abc},0)^2\frac{N_{abc}P_{abc}(q_{i+1})}{\|N_{abc}\|^2+\epsilon},
\end{eqnarray*}
which is $C^1$-continuous, and we set $\epsilon=1e^{-6}$ to avoid degeneracy. This new model only relates drag forces with the normal component of the relative velocity. Since no other constraints or conditions are imposed on $\fext{}{}$, we define $\ENG{env}=0$ for the fluid drag model.
\subsection{Controller Parametrization and Shuffle Avoidance}\label{sec:hint}
The two terms, $\ENG{phys},\ENG{env}$, cannot uniquely determine an animation. Therefore, we add two terms that model the prior knowledge in plausible character animations: controller parametrization and shuffle avoidance.

\subsubsection{Periodic and Temporal Smoothness}
\begin{figure}[t]
\begin{center}
\includegraphics[width=0.49\textwidth]{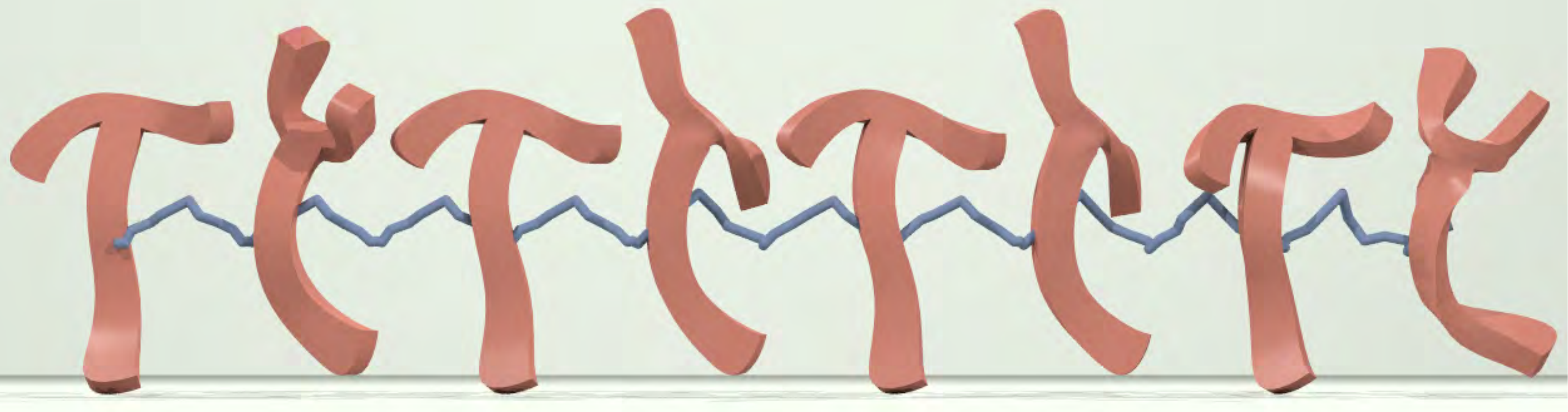}
\end{center}
\caption{\label{fig:periodic} A letter T jumping forward. With DMP regularization term $\ENG{dmp}$, its center of mass (blue) traces out a periodic trajectory.}
\end{figure}

First, we notice that for several kinds of animations, including walking, swimming, and rolling, the deformable body should move in a periodic manner. Moreover, the desired animation is temporally smooth. 
To respect this property, we use a general representation: Dynamic Movement Primitives (DMP) \cite{schaal2006dynamic} to parameterize the control inputs. DMP is a special open-loop controller parametrization that can represent many complex robotic tasks such as tennis playing and walking. DMP is capable of representing both periodic and non-periodic tasks. The latter is useful, e.g., for jumping animations. A periodic DMP controller is defined as:
\begin{eqnarray}
\label{eq:DMPP}
DMP_{p}(t,\DMPW{})=\sum_{n=1}^\mathcal{N}\alpha_n\E{exp}(\beta_n^2\E{cos}(\tau t-\mu_n)),
\end{eqnarray}
and a non-periodic DMP controller is defined as:
\begin{eqnarray}
\label{eq:DMPNP}
DMP_{np}(t,\DMPW{})=\sum_{n=1}^\mathcal{N}\alpha_n\E{exp}(-(\beta_n t-\mu_n)^2)t.
\end{eqnarray}
Note that DMP can be considered a special kind of one-input-one-output neural network using $\E{exp}()$ and $\E{cos}()$ as the activation function, where $\mathcal{N}$ is the number of neurons in each layer and the neural-net weights are $\DMPW{}\triangleq(\alpha_n,\beta_n,\mu_n,\tau)$. In practice, we need one DMP function for each component of $\fint{i}{}$ so that the total number of additional variables to be determined is $|\DMPW{}|\times|\fint{i}{}|=|\DMPW{}|\times|u|$. We denote the DMP for the $j$th component of $\fint{i}{}$ using superscript $j$. In order to guide the optimizer to look for control inputs that can be represented using DMP, we introduce an additional energy term:
\begin{eqnarray*}
\ENG{dmp}(\fint{i}{})=\frac{1}{2}\sum_{j=1}^{|\fint{i}{}|}\|\fint{i}{j}-DMP_{p/np}(i\Delta t,\DMPW{j})\|^2.
\end{eqnarray*}
In practice, we simultaneously optimize $\traj$ and $\DMPW{j}$. \changed{We also adaptively adjust the weighting of this term so that $\ENG{dmp}$ is almost zero after the iterative algorithm converges. As a result, the output of DMP function $DMP_{p/np}$ matches the required internal control forces $\fint{i}{j}$ exactly and $DMP_{p/np}$ can be used as an open-loop controller after spacetime optimization. To achieve such exact match between $\fint{i}{j}$ and $DMP_{p/np}$, we use a simple adaptive penalty method \cite{Boyd:2011:DOS:2185815.2185816}. Specifically, we use \prettyref{alg:cdmp} to adjust $\COEF{dmp}$ after every iteration of optimization. Our scheme allows the optimizer to quickly explore the space of new animations, while keeping $\|\fint{i}{j}-DMP(i\Delta t,\mathcal{W}^j)\|^2$ small. \prettyref{fig:periodic} illustrates the effect of this heuristic term.}
\setlength{\textfloatsep}{2pt}
\begin{algorithm}[ht]
\caption{Algorithm to updated $\COEF{dmp}$.}
\label{alg:cdmp}
\begin{algorithmic}[1]
\State Evaluate $A\gets\sum_{i=2}^{K-1}\ENG{eom}(q_{i-1},q_i,q_{i+1})$
\State Evaluate $B\gets\sum_{i=2}^{K-1}\|\fint{i}{j}-DMP(i\Delta t,\mathcal{W}_j)\|^2$
\If{$B > 0.1A$}\Comment Control input does not match DMP
\State $\COEF{dmp}\gets 2.1\COEF{dmp}$\Comment Enforce better match
\EndIf
\If{$B < 0.01A$}\Comment Control input matches DMP
\State $\COEF{dmp}\gets 0.5\COEF{dmp}$\Comment Allow more animation explorations
\EndIf
\end{algorithmic}
\end{algorithm}

\subsubsection{Shuffle Avoidance}
\begin{figure}[t]
\begin{center}
\scalebox{0.95}{
\includegraphics[width=0.49\textwidth]{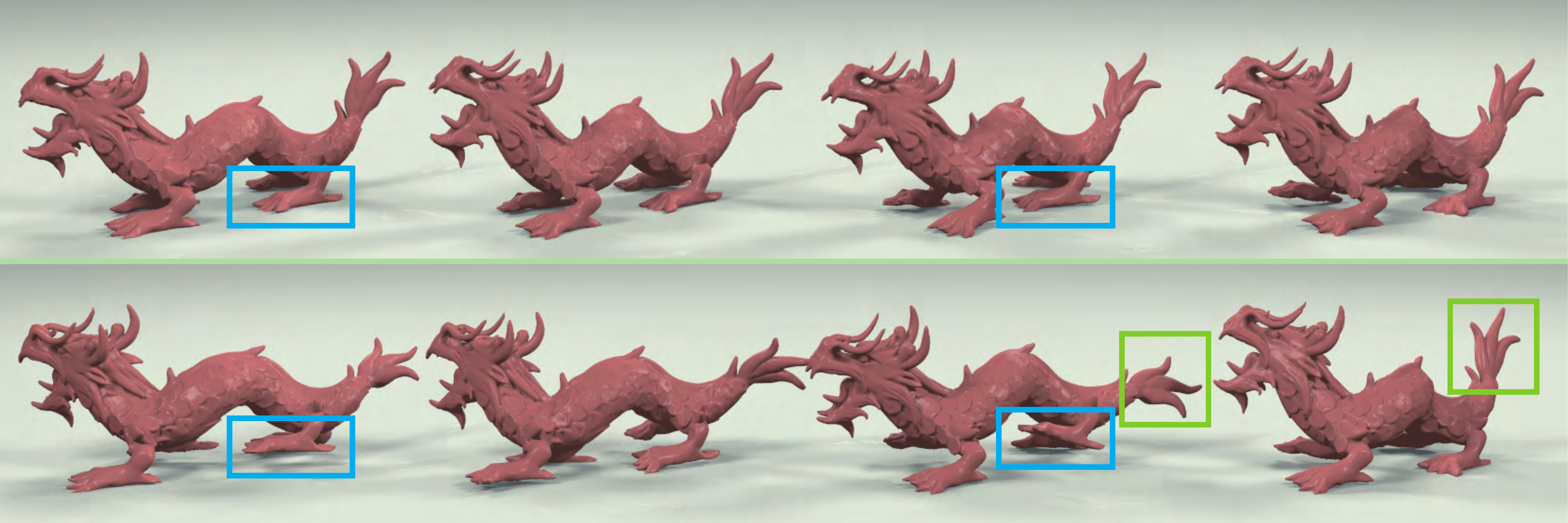}
\put(-220,75){(a)}
\put(-220,30){(b)}}
\end{center}
\caption{\label{fig:shuffle} Frames highlighting the dragon walking trajectory using our approach. In the result without $\ENG{shuffle}$ (a), the dragon's foot is always in close proximity to the floor plane (blue). The artifact is mitigated using $\ENG{shuffle}$ (b). Therefore, we observe more secondary dynamics in (b), e.g., around the tail (green).}
\end{figure}

As observed in \cite{Mordatch:2013:AHL:2508363.2508365}, another artifact due to the lack of internal actuation structure is the shuffling movement across the contact manifold. This means that the contact points are always in close proximity to the solid boundary. 
To mitigate this artifact, we introduce an additional hint term $\ENG{shuffle}$ defined as:
\begin{eqnarray*}
\ENG{shuffle}(q_i,q_{i-1})=
\sum_{j=1}^V\|(v^j(q_i)-v^j(q_{i-1}))_\parallel\|^2\E{exp}(-\gamma\E{dist}(v^j(q_i))),
\end{eqnarray*}
where $\gamma$ is the distance attenuation coefficient. For each vertex $v^j$, we penalize its tangential velocity attenuated by its distance from static obstacles. 
In this way, the shuffling artifact is removed by asking a walker to lift its legs to move forward. The effect of this hint term is illustrated in \prettyref{fig:shuffle}. We combine the above two hints and Tikhonov regularization, giving:
\begin{eqnarray*}
\resizebox{.47\textwidth}{!}{$
\ENG{hint}(\traj)=\sum_{i=2}^{K-1}\frac{\COEF{reg}}{2}\|\fint{i}{}\|^2+
\COEF{shuffle}\ENG{shuffle}(q_i,q_{i-1})+\COEF{dmp}{\ENG{dmp}}(\fint{i}{}).
$}
\end{eqnarray*}
\section{SpaceTime Optimization}\label{sec:optimization}
\setlength{\textfloatsep}{5pt}
\begin{algorithm}[ht]
\caption{The hybrid optimization algorithm.}
\label{alg:opt}
\begin{algorithmic}[1]
\LineComment \changed{Setup multiple initial guesses for DMP periods}
\State \changed{$2\pi/\tau\gets0.2,0.4,\cdots,5(s)$}
\For{iteration=$0,\cdots$}
\LineComment Evaluate function values and gradients
\State Evaluate $E(\traj)$, $\FPP{E(\traj)}{\traj}$
\State Evaluate an approximation to $\FPPT{E(\traj)}{\traj}$ denoted as $H$

\LineComment Update trajectory, see \cite{lourakis2005brief} for more details
\Do
  \label{ln:linsol}
  \State $\traj^{new}\gets \traj-(H+d\E{I})^{-1}\FPP{E(\traj)}{\traj}$
  \If{$E(\traj^{new}) > E(\traj)$}
    \State increase $d$
  \Else
    \State decrease $d$
  \EndIf
\doWhile{$E(\traj^{new}) > E(\traj)$}
\State $\traj\gets \traj^{new}$

\LineComment Update contact forces
\If{Using frictional contact model}
  \For{$i=2,\cdots,K-1$ in parallel}
    \State Update $\fext{i}{}$ using \cite{5979814}
  \EndFor
\EndIf

\LineComment \changed{Update DMP weights from each initial guess}
\If{$\DMPW{}$ is not updated in the last 10 iterations}
  \For{\changed{Each initial guess}}
  \label{ln:multiLBFGS}
    \State Update DMP weights $\DMPW{}$ using 1000
    \State iterations of LBFGS \cite{Liu1989}
  \EndFor
  \State \changed{Choose $\DMPW{}$ leading to smallest $\ENG{dmp}$}
\EndIf

\If{Using frictional contact model}
  \For{all $1\leq i\leq K$ and $1\leq j \leq V$}
  \label{ln:cio}
    \If{\changed{$\E{dist}(v^j(q_i)) > \epsilon_1$ and $\|\fext{i}{j}\| < \epsilon_2 \fmax{i,j}{\|\fext{i}{j}\|}$}}
      \State Exclude $v^j(q_i)$ from $\ENG{env,shuffle}$
    \Else
      \State Include $v^j(q_i)$ in $\ENG{env,shuffle}$
    \EndIf
  \EndFor
\EndIf
\EndFor
\end{algorithmic}
\end{algorithm}
In this section, we present our efficient, hybrid optimizer to minimize the objective function:
\begin{eqnarray}
\label{eq:prob}
\argmin{\traj,\fint{}{},\fext{}{},\DMPW{}}E(\traj,\fint{}{},\fext{}{},\DMPW{}),
\end{eqnarray}
where different subproblem solvers are used for minimizing with respect to each of the 4 free variables: $\traj,\fint{}{},\fext{}{},\DMPW{}$. As a special case, $\fext{}{}$ is not a free variable for swimming animations using our fluid drag model. Without loss of generality, we consider \prettyref{eq:prob} for presentation. Since the objective function is $C^1$-continuous, our first attempt was to use an off-the-shelf implementation of the LBFGS algorithm \cite{Liu1989}. However, we found out that even for small problems, having very small $|u|$ and $K$, it takes a large number of iterations to converge. Instead, we present a novel hybrid optimization algorithm that converges in much fewer iterations.

\subsection{Hybrid Optimizer}
\begin{table}
\centering
\setlength{\tabcolsep}{20pt}
\begin{tabular}{lcc}
\toprule
Example & LBFGS(s) & Hybrid(s)	\\
\midrule
2D Crawling & 1534 & 18    \\
2D Rolling  & 823  & 12   \\
\hline \\
\end{tabular}
\caption{\label{table:perf} Performance of LBFGS and our hybrid solver on two examples: 2D worm crawling and 2D ball rolling. Our approach is significantly faster.}
\end{table}
To accelerate the rate of convergence, we first notice that $\fint{}{}$ appears only in $\ENG{eom}$, and $\ENG{dmp}$ as a quadratic function. Therefore, we can solve for $\fint{}{}$ analytically and eliminate it. We further observe that the other three sets of variables ($\fint{}{},\fext{}{},\DMPW{}$) appear in the objective function with special structures. The DMP weight vector $\DMPW{}$ appears only in $\ENG{dmp}$ and optimizing $\DMPW{}$ amounts to a small neural-network training problem for which LBFGS is proven to be very effective. The external force $\fext{}{}$ is a quadratic function in both $\ENG{eom}$ and $\ENG{env}$, and $\fext{i}{}$ for each timestep $i$ is separable and can be solved in parallel. Together with constraint \prettyref{eq:cone}, finding the optimal $\fext{}{}$ amounts to solving a QCQP problem, for which special solvers are known. For example, we use a primal interior-point method \cite{5979814}. We found that solving QCQP is faster than solving QP with a linearized frictional model because it requires fewer constraints and makes use of coherence in the intermediate solutions between the consecutive iterations by allowing warm-starting. We can update these variables $\DMPW{}$, $\fext{}{}$, and $\traj$ in an alternate manner. Finally, for trajectory $\traj$ itself, LBFGS can still be used, but we found that LBFGS does not use gradient information effectively. A large number of gradient evaluations are performed inside the line-search scheme and LBFGS usually chooses a conservative step size. Therefore, we choose the Levenberg-Marquardt (LM) method for updating $\traj$. \changed{The outline of our method is is given in \prettyref{alg:opt}, and we provide the low-level details in \refSuppOptimizer.} \prettyref{table:perf} shows a comparison between our solver and LBFGS on two small 2D problems.

\subsection{Efficient Function Evaluation}
The costliest step in our algorithm is the evaluation of the function values, including the gradient and approximate hessian. We combine several techniques to accelerate these evaluations. First, we notice that $v^j(q_i)$, the recovered vertex $j$'s Euclidean coordinates from reduced representation at timestep $i$, appears in almost every objective term. Moreover, these values are independent of each other. Therefore, we can compute and store $v^j(q_i)$, $\FPP{v^j}{q_i}$, $\FPPT{v^j}{q_i}$ for all $1\leq i\leq K$ and $1\leq j\leq V$ in parallel, before each evaluation. We provide some hints for computing the second derivatives $\FPPT{v^j}{q_i}$ in \refSuppDeriv. The overhead of these computations is independent of the number of vertices $V$. We also utilize this information to assemble the hessian. This assembly step can be a computational bottleneck because we have to evaluate the summations over all the vertices that appear in the physics violation term $\ENG{eom}$, in the collision avoidance term $\ENG{coll,self}$, in the environmental force term $\ENG{env}$, and finally in the shuffle avoidance term $\ENG{shuffle}$.

\subsubsection{Accelerating the Assembly of $\ENG{eom}$}
\begin{figure}[t]
\begin{center}
\scalebox{0.95}{
\includegraphics[width=0.49\textwidth]{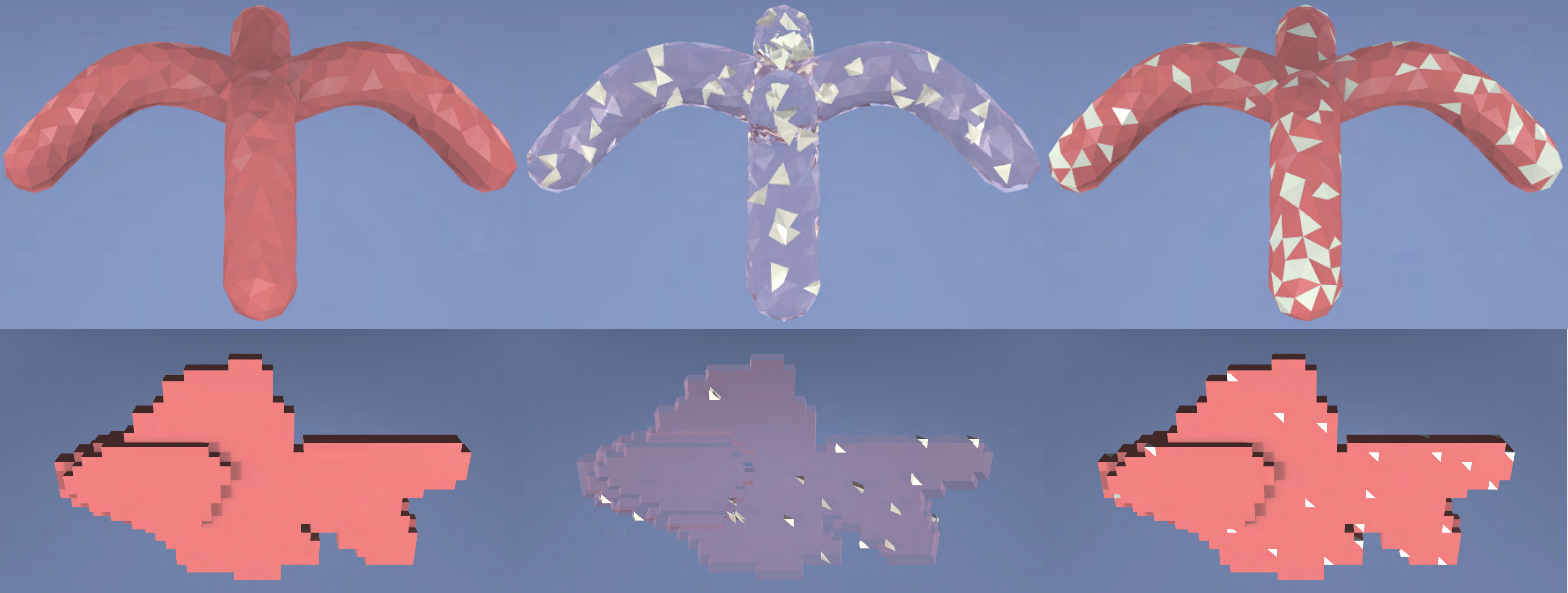}
\put(-190,40){(a)}
\put(-105,40){(b)}
\put(-20 ,40){(c)}}
\end{center}
\caption{\label{fig:cubature} \changed{For the spider (top) and fish (bottom) models (a), we visualize the kinetic cubatures (b) and surface patch cubatures (c). In both cases, only a small fraction of elements need to be considered for the summation. This fraction is $12\%$ for the spider and $0.7\%$ for the fish model.}}
\end{figure}

\cite{An:2008:OCE:1457515.1409118,Barbic:2005:RSI:1186822.1073300} have addressed the problem of accelerating the assembly of $\ENG{eom}$. Specifically, summation over all vertices appears in two places of $\ENG{eom}$ highlighted below:
\begin{small}
\begin{eqnarray*}
\colorboxed{blue}{\FPP{\bar{q}}{q}^T(q_{i+1})MA(q_{i+1})}+
\FPP{\left[P-\fint{i}{}^Tu-\colorboxed{red}{\fext{i}{}^T\bar{q}}\right]}{q}(q_{i+1}).
\end{eqnarray*}
\end{small}
We use the cubature approximation \cite{An:2008:OCE:1457515.1409118} to accelerate these two terms. The blue part above corresponds to the kinetic cubature used in \cite{Pan:2015:SDS:2816795.2818090}; see \refSuppKinetic for more details. The red part above corresponds to the fluid drag force, which is a summation over all the surface patches. Cubature approximation essentially assumes that:
\begin{eqnarray}
\label{eq:cubature}
&&\fext{}{T}(q_{i+1},q_i)\bar{q}(q_{i+1})    \\
&=&\COEF{drag}\sum_{(v^a,v^b,v^c)}\E{max}(N_{abc}^TU_{abc},0)^2\frac{N_{abc}P_{abc}(q_{i+1})}{\|N_{abc}\|^2+\epsilon} \nonumber  \\
&\approx&\sum_{(v^a,v^b,v^c)\in\mathcal{T}}C^{abc}\E{max}(N_{abc}^TU_{abc},0)^2\frac{N_{abc}P_{abc}(q_{i+1})}{\|N_{abc}\|^2+\epsilon}, \nonumber
\end{eqnarray}
i.e., the sum over all surface patches can be approximated using the weighted sum of a selected set of surface patches $\mathcal{T}$. The set $\mathcal{T}$ and weights $C^{abc}$ are computed via dictionary learning. As illustrated in \prettyref{fig:cubature}, this greatly reduces the computational overhead.

\subsubsection{Accelerating the Assembly of $\ENG{coll,self}$}
For collision avoidance terms, only very few vertices will contribute non-zero values to the objective function. Therefore, we use a bounding volume hierarchy \cite{James:2004:BOC:1186562.1015735} to update the non-zero terms. This data-structure can be updated solely using reduced representation $|q|$, and the update for different timesteps can be performed in parallel.

\subsubsection{Accelerating the Assembly of $\ENG{env,shuffle}$}
In the previous section, we used cubature approximation to accelerate the fluid drag forces. For frictional contact forces, however, all the vertices in close proximity to the static obstacles will contribute non-zero values to $\ENG{env}$ and $\ENG{shuffle}$. Since these vertices cannot be determined during the precomputation stage, we dynamically update them. Specifically, we remove vertex $v^j(q_i)$ from $\ENG{env}$ if \changed{$\E{dist}(v^j(q_i)) > \epsilon_1$ and $\|\fext{i}{j}\| < \epsilon_2 \fmax{i,j}{\|\fext{i}{j}\|}$}. After $\ENG{env}$ is updated, we update $\ENG{shuffle}$ accordingly, since $\ENG{shuffle}$ is also very small for vertices that are far from the static obstacles. These updates can be accelerated using a bounding volume hierarchy.

\begin{changedBlk}
\subsection{Robustness to Suboptimal Solutions}\label{sec:localSol}
It is well-known that spacetime optimization is prone to bad local minima leading to suboptimal solutions, except for simple cases \cite{Barbic:2012:IED:2185520.2185566}. In our algorithm, there are two energy terms that can result in the computation of bad local minima. One is the contact integrity term, $\ENG{env}$, which models the non-smoothness of frictional contacts. The other one is $\ENG{dmp}$, which models the trajectory smoothness and periodic movements. 

In terms of $\ENG{dmp}$, previous methods~\cite{schaal2006dynamic,10.3389/fncom.2013.00138} use sampling-based methods to search for the global optimum. Since we only use a gradient-based local optimizer, $\ENG{dmp}$ could result in the computation of a bad local minima. Indeed, we found that our optimizer can have difficulty in terms of finding good DMP parameters $\mathcal{W}$. At a local minima, several DMP neurons usually have same values of $(\alpha_n,\beta_n,\mu_n)$, values in \prettyref{eq:DMPP} or \prettyref{eq:DMPNP}, meaning that we are wasting parameters. In addition, we found that the period parameter $\tau$ can get stuck in a local minima very close to our initial guess. In this section, we introduce some simple modifications to overcome these problems.
\begin{wrapfigure}{r}{4.5cm}
\begin{center}
\includegraphics[width=0.23\textwidth]{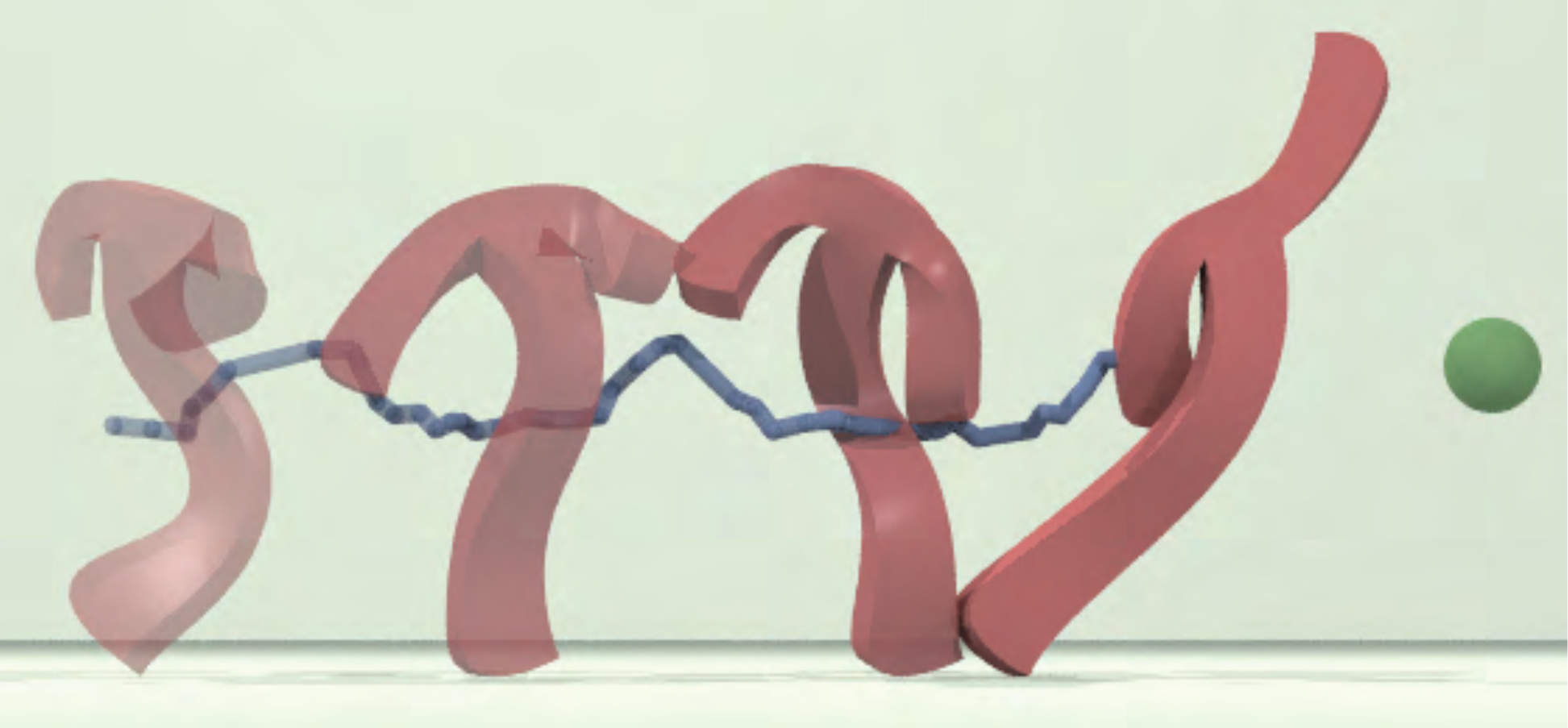}
\end{center}
\caption{\label{fig:failure} \changed{When user sets a target point (green) too far away ($5$ meters to the right) and uses very few timesteps ($20$ in this case), the Letter T chooses to lean itself too much to recover from falling down.}}
\end{wrapfigure}

We first initialize the phase shift uniformly in the phase space, i.e., $\mu_n=2\pi n/N$, and we initialize $\alpha_n,\beta_n$ to very small random values. To avoid the period parameter $\tau$ falling into a bad local minima, we use multiple initial guess for $\tau$ and run an LBFGS optimization from each initial guess in \prettyref{ln:multiLBFGS} of \prettyref{alg:opt}. In our experiments, we set $2\pi/\tau=0.2,0.4,\cdots,5(s)$ and run LBFGS 25 times very 10 iterations to avoid bad local minima. After that, we get 25 candidate DMP parameters, $\mathcal{W}$, and we choose the candidate leading to the smallest $\ENG{dmp}$. Such multiple LBFGS optimizations will result in additional computational overhead during the first few iterations of optimization. As the optimizer gets closer to a local minima, $\tau$ will converge to a same local minima for several candidates of DMP parameters, and we can merge these candidates into one. In addition, if a certain candidate is never chosen as the best during the last 100 iterations, we remove this candidate from further consideration. In practice, we have only $2-3$ remaining candidates after $500$ iterations. 

The approach highlighted above greatly increases the chances that our optimization algorithm computes a good local minima without significant computational overhead. This is because periodic DMP formulation (\prettyref{eq:DMPP}) is guiding the whole trajectory to follow a same gait. When our optimization algorithm finds a useful gait, this information is quickly encoded into the DMP controller and reused to compute the entire trajectory using the $\ENG{dmp}$ formulation. In order to highlight this feature, we show two swimming trajectories computed using our optimization algorithm. In order to compute the trajectory shown in \prettyref{fig:randomInit} (a), we initialize the spider pose to $u=c=w=0$ at all timesteps. While to generate \prettyref{fig:randomInit} (b), we initialize the spider to a different random pose at every timestep. Moreover, the convergence history of these two optimization schemes are plotted in \prettyref{fig:convHis}. Our optimizer converges to two different but almost equally effective swimming gaits with very small objective function values. This means that although there are numerous local minima, most of them leads to plausible animations. However, bad local minima can still happen especially in contact-rich animations and we illustrate one such failure case in \prettyref{fig:failure}.
\begin{figure}[ht]
\begin{center}
\includegraphics[width=0.49\textwidth]{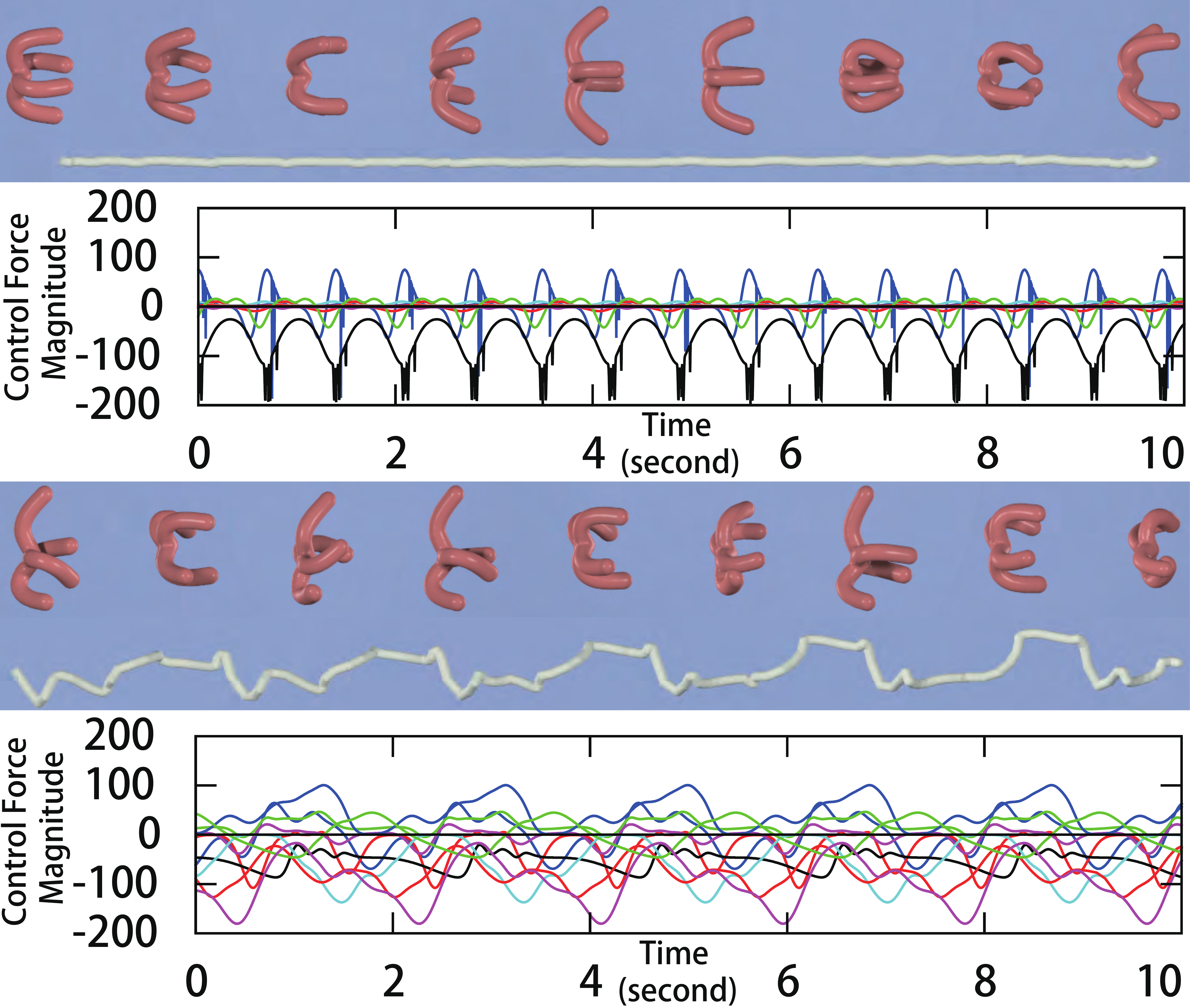}
\put(-25,200){(a)}
\put(-25,100){(b)}
\put(-15,160){(c)}
\put(-15, 49){(d)}
\end{center}
\caption{\label{fig:randomInit} \changed{We show two swimming trajectories optimized using static initialization (a) and random initialization (b). For both trajectories, we plot the locus of the deformable body's center of mass (white curve) and the magnitude of control forces in (c,d). The goal is to move 5 meters to the left after 10 seconds. Our optimizer finds two different but almost equally effective gaits.}}
\end{figure}
\begin{figure}[ht]
\begin{center}
\includegraphics[width=0.49\textwidth]{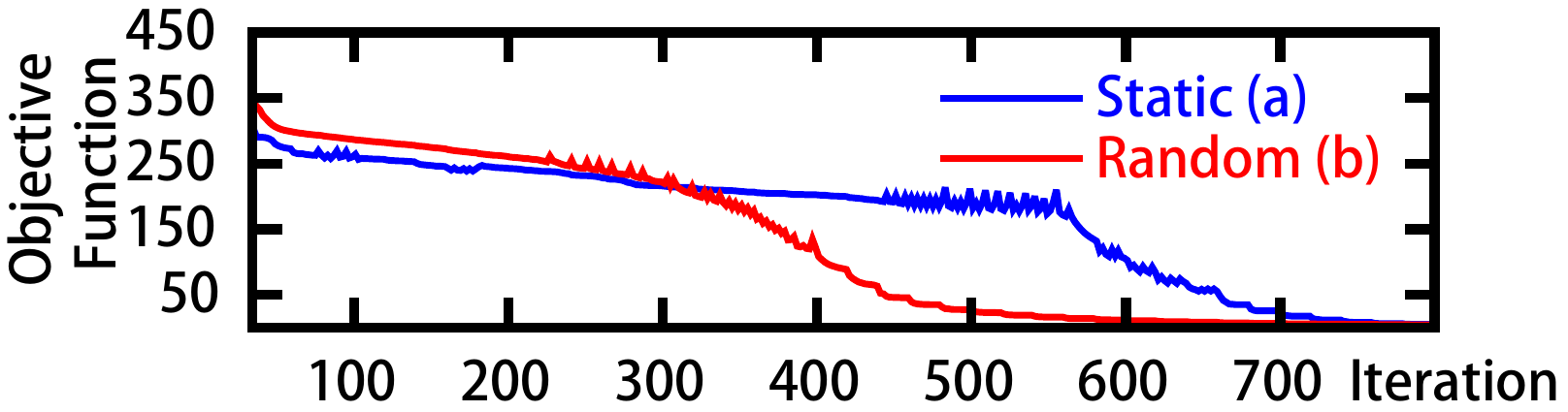}
\end{center}
\caption{\label{fig:convHis} \changed{We plot the convergence history using static initialization (a) and random initialization (b). Both optimizations reduce the objective function to less than $1\%$ of the original value. This plot shows that many local minima of our objective function leads to plausible animations. There are some jittering during the optimization. This is because the adaptive penalty method (\prettyref{alg:cdmp}) is adjusting $\COEF{dmp}$.}}
\end{figure}
\end{changedBlk}
\section{Results}\label{sec:results}
In this section, we highlight the results on complex benchmarks.

\begin{table}
\centering
\setlength{\tabcolsep}{20pt}
\begin{tabular}{ll}
\toprule
Name & Value \\
\midrule
$\COEF{coll}$ & $1e^2$ \\
$\COEF{self}$ & $1e^2$ \\
$\COEF{env}$ & $1e^1/l^2$ \\
$\COEF{drag}$ & $1e^3$ \\
$\COEF{reg}$ & $1e^{-3}$ \\
$\COEF{shffle}$ & $1e^{-1}$ \\
$\COEF{dmp}$ & dynamic \\
$\mu$ & $0.7$ \\
$\gamma$ & $\E{log}(10)/l$ \\
\changed{$\epsilon_1$} & $3l$ \\
\changed{$\epsilon_2$} & $0.01$ \\
$\Delta t$ & $0.05s$ \\
Young's modulus & $1e^5Pa$ \\
Poisson's ratio & $0.48$ \\
Mass density $\rho$ & $1kg/m^3$ \\
Gravity & $9.81m/s^2$ \\
$\mathcal{N}$ & $5$ \\
\hline \\
\end{tabular}
\caption{\label{table:param} Parameters.}
\end{table}
\TE{Parameter Choices}: We use an identical set of parameters listed in \prettyref{table:param} for all the benchmarks. The coefficient of the physics violation term is $1$. Some parameters are related to $l$, which is the average element size. If a deformable body has volume $vol$ and is discretized using $P$ FEM elements, then $l=(vol/P)^{1/3}$. An exception is the coefficient for $\ENG{dmp}$, which is adaptively adjusted within the optimization algorithm.

\TE{Benchmarks}: We implemented our method in C++ and tested it on many benchmarks using a desktop machine with dual E5-2670 12-core CPU 2.1GHz and 12GB of memory. Given only a volumetric mesh and a definition of the environment, we first precompute the reduced dynamic model using \cite{Pan:2015:SDS:2816795.2818090}. We also precompute the surface cubatures to approximate the fluid drag forces. We use OpenMP to parallelize the function and gradient evaluations and run at most 10000 iterations of optimizations or stop early, if the relative error of $\|\FPP{E(\traj)}{\traj}\|$ is smaller than $1e^{-3}$. The setup and computational cost in each benchmark is summarized in \prettyref{table:bench} and analyzed below.
\begin{table}[ht]
\setlength{\tabcolsep}{2pt}
\begin{center}
\scalebox{0.77}{
\begin{tabular}{lcccccc}
\toprule
Example                                           & $V/P$     & $|u|$        & Pre./PreSF.(min) & K/$\#\traj$ & Opt.(hr) & App.	\\
\midrule
Fish Swimming (\shortref{fig:swim}{a})            & 2118/7812 & 5  & 0.8/0.1 & 200/3 & 1.7 & \textcolor{black}{DMP}  \\
Spider Swimming (\shortref{fig:swim}{b})          & 1054/4033 & 10 & 1.2/0.3 & 200/3 & 2.5 & \textcolor{black}{DMP}  \\
\begin{tabular}{@{}l@{}}\changed{Spider Swimming} \\ \changed{Reduced StVK (\shortref{fig:STVKLetterT}{a})}\end{tabular}
                                                  & 1054/4033 & 65 & 4.6/0.3 & 200/1 & 3.9 & None   \\
Spider Walking (\shortref{fig:walkGeom}{})        & 1054/4033 & 10 & 1.2/    & 200/4 & 5.2 & \textcolor{black}{FB}   \\
Dragon Walking (\shortref{fig:shuffle}{})         & 929/1854  & 10 & 1.3/    & 200/1 & 2.2 & None   \\
Letter T Walking (\shortref{fig:trackT}{a})       & 1523/3042 & 15 & 1.1/    & 200/4 & 4.5 & \textcolor{black}{FB}   \\
\begin{tabular}{@{}l@{}}\changed{Letter T Walking} \\ \changed{Reduced StVK (\shortref{fig:STVKLetterT}{b})}\end{tabular}
                                                  & 1523/3042 & 65 & 4.2/    & 200/1 & 3.1 & None   \\
Beam Jumping (\shortref{fig:jumpHeight}{})        & 1024/640  & 10 & 1.1/    & 100/1 & 1.1 & None   \\
Cross Rolling (\shortref{fig:roll}{})             & 623/1499  & 10 & 1.3/    & 200/1 & 2.1 & None   \\
Dinosaur Walking (\shortref{fig:dinoWalk}{})      & 1493/5249 & 15 & 0.9/    & 200/1 & 1.9 & None   \\
\bottomrule
\end{tabular}}
\end{center}
\caption{\label{table:bench} Benchmark setup and computational overhead. From left to right, number of vertices $V$/number of FEM elements $P$, DOFs of local deformation $|u|$, precomputation time for building reduced dynamic model/computing surface patch cubatures, number of frames/number of trajectories, time spent on optimization, and the supported application: \textcolor{black}{DMP} means we use DMP as open-loop controller to drive forward simulation, \textcolor{black}{FB} means that we use feedback controller to track the animation (both of these are realtime).}
\end{table}

\begin{figure}[t]
\begin{center}
\scalebox{0.95}{
\includegraphics[width=0.49\textwidth]{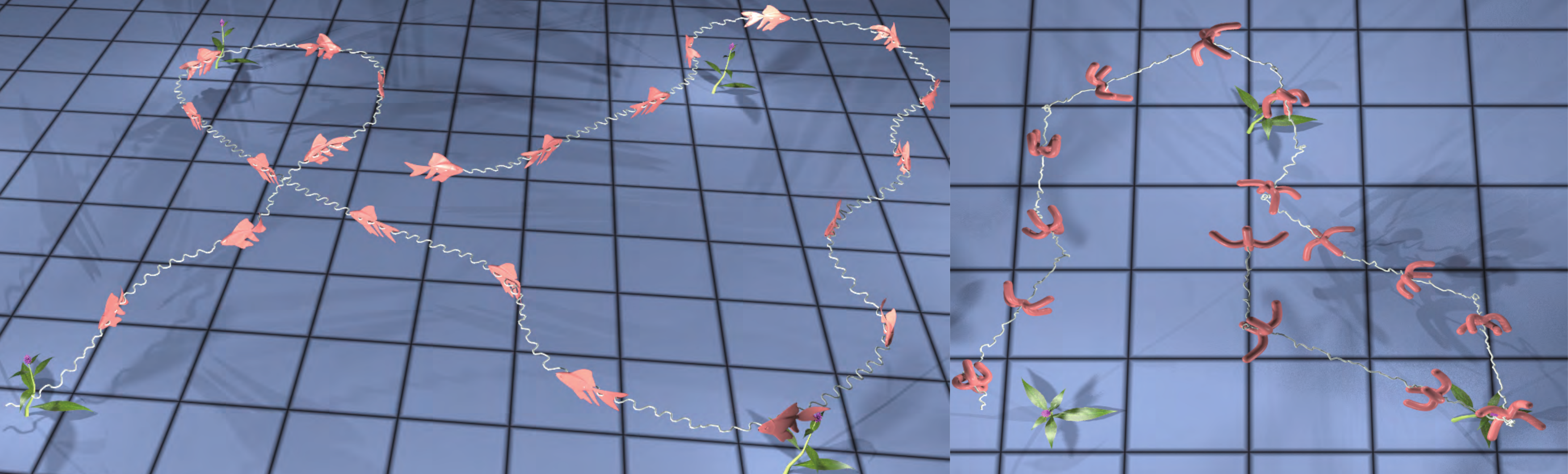}
\put(-220,5){(a)}
\put(-70 ,5){(b)}}
\end{center}
\caption{\label{fig:swim} We use RRT* to navigate physical swimming characters, the fish (a) and the spider (b), to look for food plants (green). The white line is the locus of the deformable body's center of mass computed using RRT*.}
\end{figure}

\TE{Fish Swimming}: Fishes have the simplest deformable bodies and can be used for testing the performance of our method. As illustrated in \prettyref{fig:Teaser}, a fish swims by simply swinging its body, so we use a reduced configuration space of small DOFs: $|u|=5$, i.e., $|q|=11$. Under this setting, we command the fish to swim straight forward in a gravityless environment using the following objective:
\begin{small}
\begin{eqnarray}
\label{eq:forward}
\ENGS{obj}{move}(\traj)=\COEFS{obj}{move}\sum_{k=2}^K\|c_{k+1}-c_{k}-\E{exp}(w_{k+1})v_c\Delta t\|^2/2,
\end{eqnarray}
\end{small}
where $\E{exp}(w_{k+1})$ transforms the velocity to a global frame of reference and $v_c$ is the target swimming speed in a local frame of reference. In addition, we add a balance energy to encourage fixed orientation:
\begin{eqnarray}
\label{eq:balance}
\ENGS{obj}{bal}(\traj,d)=\COEFS{obj}{bal}\sum_{k=2}^K\|\E{exp}(w_{k})d-d\|^2/2,
\end{eqnarray}
where $d$ is the balance direction. \changed{Here we use $d=g$, the unit gravitational direction.} We can even navigate the fish to an arbitrary 3D point by optimizing 3 trajectories: swimming forward, swimming left, and swimming right. For swimming left and right, we add the following objective functions in addition to \prettyref{eq:forward}:
\begin{small}
\begin{eqnarray}
\label{eq:turn}
\ENGS{obj}{turn}(\traj,d)=\COEFS{obj}{turn}\sum_{k=2}^K\|\exp(w_{k+1})-\E{exp}(\theta d\Delta t)\exp(w_k)\|^2/2,
\end{eqnarray}
\end{small}
where $\theta d$ is the target rotating speed, we use $d=g$ again. After the optimization, the DMP function can be used as an open-loop controller to generate controlled forward simulations at real-time framerate. In \prettyref{fig:swim} (a), we wrap our forward simulator into a sampling-based motion planner, RRT* \cite{doi:10.1177/0278364911406761}, to navigate the fish to look for food plants.

\TE{Spider Swimming}: We also evaluated our approach on a more complex model: a four-legged spider. More degrees of freedom are used to allow each leg to move independently, so we use more DOFs: $|u|=10$, i.e., $|q|=16$. Again, we optimize to generate 3 trajectories with the same $\tau$ and $\mu_n$ for all trajectories. However, for the first trajectory we set $v_c=1,\theta=0$, and for the other two we set $v_c=0,\theta=\pm1$ so that the spider cannot turn itself around while swimming forward. This gives very different gaits for turning and swimming forward. We again use DMP to drive the realtime forward simulator and wrap it into a motion planner, as illustrated in \prettyref{fig:swim} (b).

\begin{figure}[t]
\begin{center}
\scalebox{0.95}{
\includegraphics[width=0.49\textwidth]{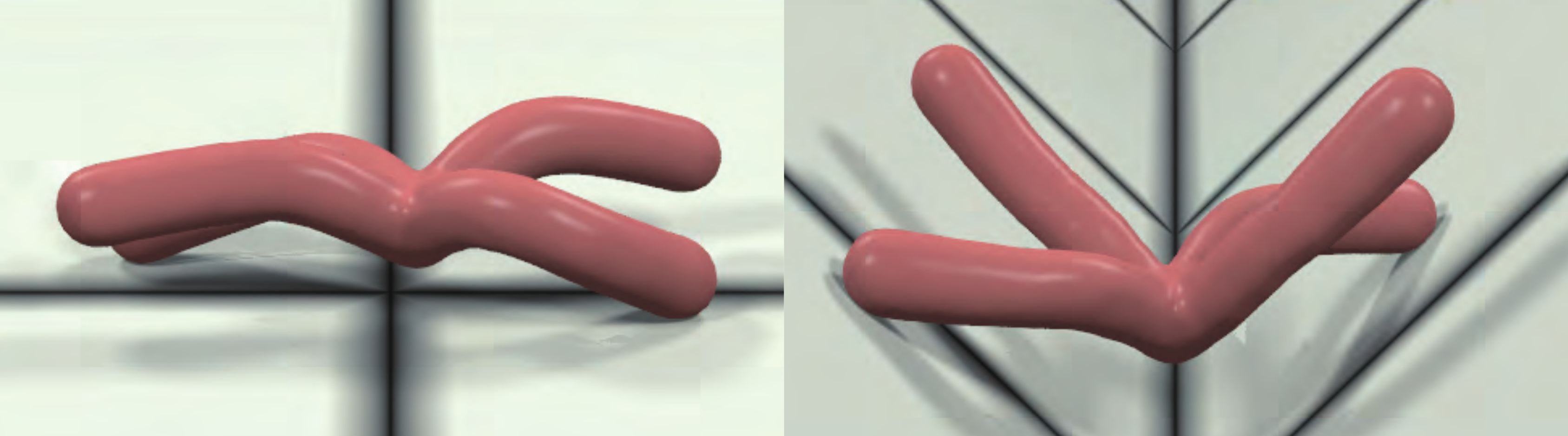}
\put(-145,5){(a)}
\put(-15 ,5){(b)}}
\end{center}
\caption{\label{fig:walkGeom} The spider walking on planar ground (a) and V-shaped ground (b).}
\end{figure}

\TE{Spider Walking}: To analyze the walking animation, we use the same spider model and objective \prettyref{eq:forward} but replace the fluid drag force model with the frictional contact force model. However, we observe that this optimization takes approximately twice as many iterations to converge due to the contact-integrity term $\ENG{env}$ and the shuffle avoidance term $\ENG{shuffle}$. In \prettyref{fig:walkGeom}, we illustrate the walking gaits for two kinds of environments.

Similar to swimming, we allow a user to navigate the spider on the ground by optimizing 4 trajectories: walking left, right, backward, and forward, where the objective function is \prettyref{eq:turn} with corresponding $v_c$. We then use a feedback controller similar to \cite{Tan:2012:SBL:2185520.2185522} to drive forward simulator. Specifically, we optimize $E(\traj)$ over one timestep ($K=3$) with the objective function:
\begin{eqnarray}
\label{eq:objTrack}
\ENGS{obj}{track}(\traj)=\COEFS{obj}{track}(\bar{q}(q_2)-\bar{q}(q^*))^TM(\bar{q}(q_2)-\bar{q}(q^*))/2,
\end{eqnarray}
where $q_k^*$ is the configuration of the tracked trajectory. Due to the efficiency of reduced representation, such short-horizon optimization can be solved at realtime framerates. 

\TE{Letter T Walking}: A more challenging example is Letter T walking, as illustrated in \prettyref{fig:periodic}. This model has no static stability, so it must keep jumping to move around. Again, we first optimize 4 trajectories and then track these trajectories at realtime to navigate the character.

\begin{figure}[t]
\begin{center}
\includegraphics[width=0.45\textwidth]{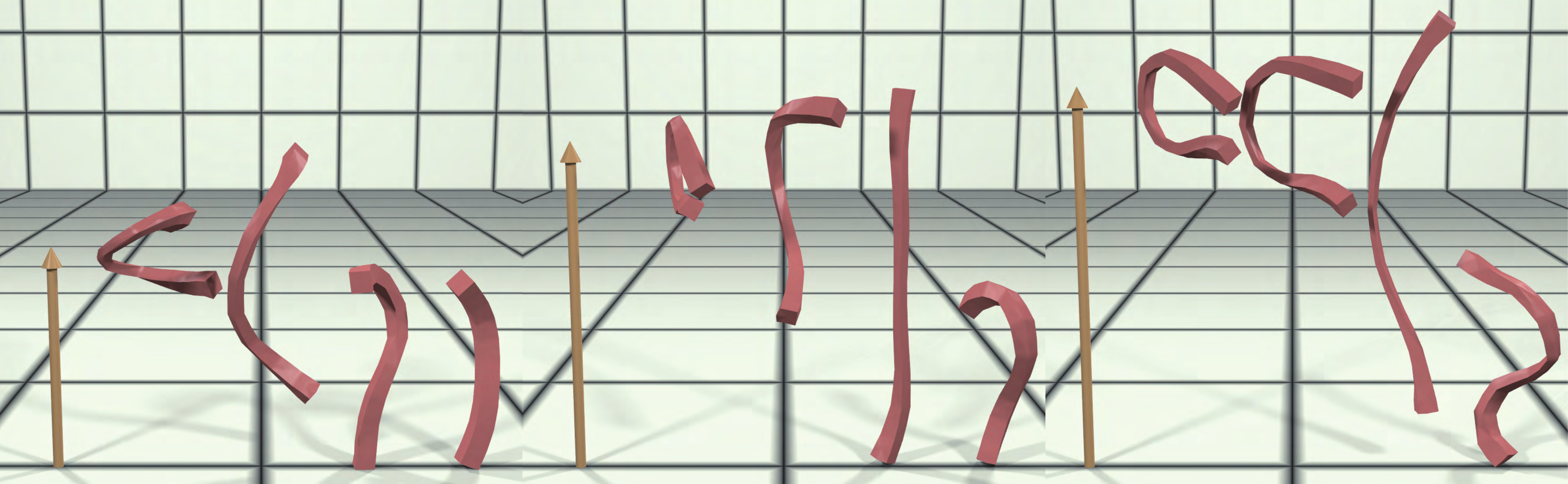}
\put(-220,5){(a)}
\put(-145,5){(b)}
\put(- 70,5){(c)}
\end{center}
\caption{\label{fig:jumpHeight} Different frames during a beam jumping with different target altitudes (yellow arrow); (a): $h=2$, (b): $h=3$, and (c): $h=4$.}
\end{figure}

\begin{figure}[t]
\begin{center}
\includegraphics[width=0.45\textwidth]{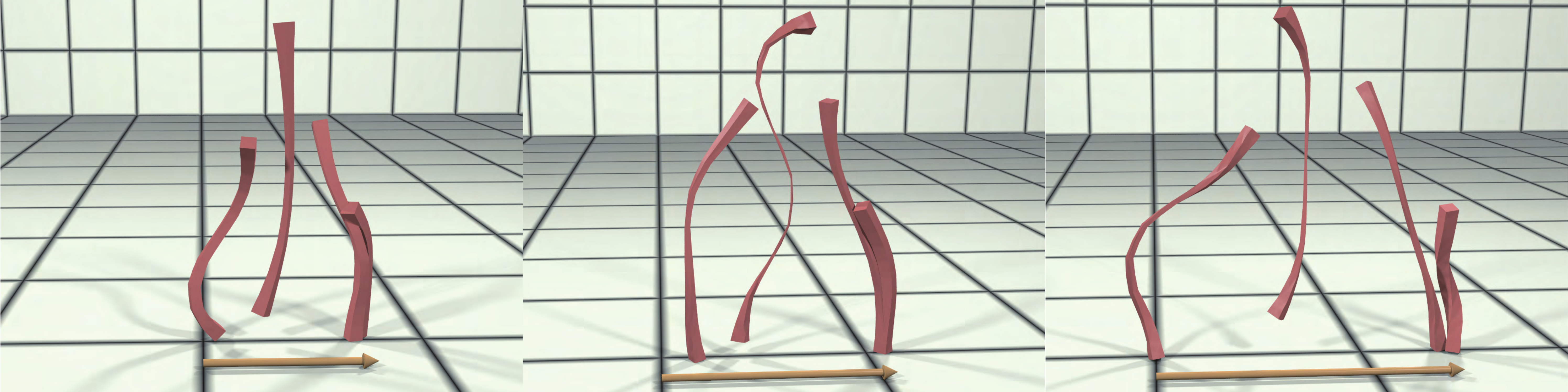}
\put(-220,5){(a)}
\put(-145,5){(b)}
\put(- 70,5){(c)}
\end{center}
\caption{\label{fig:jumpWalk} Different frames during a beam jumping forward with different target distance (yellow arrow); (a): $\|v_c\|=2$, (b): $\|v_c\|=2.5$, and (c): $\|v_c\|=3$ (the target altitude $h=3$).}
\end{figure}

\TE{Beam Jumping}: Jumping is an essential component in many animations. To generate these animations, we use the following objective function:
\begin{eqnarray}
\label{eq:jump}
&&\ENGS{obj}{jump}(\traj)=\COEFS{obj}{jump}\|g^Tc_{K/2}-h\|^2/2+    \\
&&\COEFS{obj}{jump}\|(\E{I}-gg^T)(c_{K/2}-c_{K/2-1}-\E{exp}(w_{K/2})v_c\Delta t)\|^2/2 \nonumber,
\end{eqnarray}
where the first term specifies the target altitude and the second term specifies the target horizontal velocity $v_c$ so that the character can jump forward. Using different $h$ and $v_c$, we generate a series of results in \prettyref{fig:jumpHeight} and \prettyref{fig:jumpWalk} for a small beam, where the beam exhibits huge and varied deformations.

\begin{figure}[t]
\begin{center}
\scalebox{0.96}{
\includegraphics[width=0.48\textwidth]{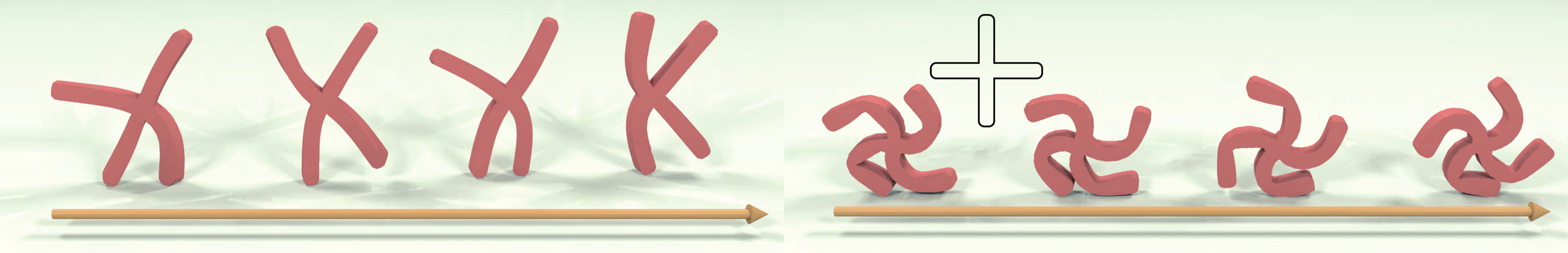}
\put(-88 ,32){(a)}
\put(-183,10){(b)}
\put(-65 ,10){(c)}}
\end{center}
\caption{\label{fig:roll} \changed{(a) Shows the rest pose (black) of an X-shaped deformable body; (b) Shows frames generated using 3D bases; (c) Highlights the frames of a rolling animation generated with 2D bases. Our optimizer deforms it into a swastika for energy-efficient rolling.}}
\end{figure}

\TE{Cross Rolling}: As illustrated in \prettyref{fig:roll} (b), we generate a rolling animation for a cross-shaped deformable body using the objective function \prettyref{eq:balance} and \prettyref{eq:turn}. In addition, we notice that the deformable body in this example is close to planar. Therefore, we generated a second animation by restricting all the deformations to the 2D plane. As shown in \prettyref{fig:roll} (c), the deformable body exhibits very different gaits in this case. \changed{Also, this result shows that our hybrid optimization algorithm can automatically compute a very different mean pose (a swastika) from the rest pose (an X) in order to perform the locomotion task in an energy-efficient manner.}
\begin{figure}[t]
\begin{center}
\scalebox{0.95}{
\includegraphics[width=0.49\textwidth]{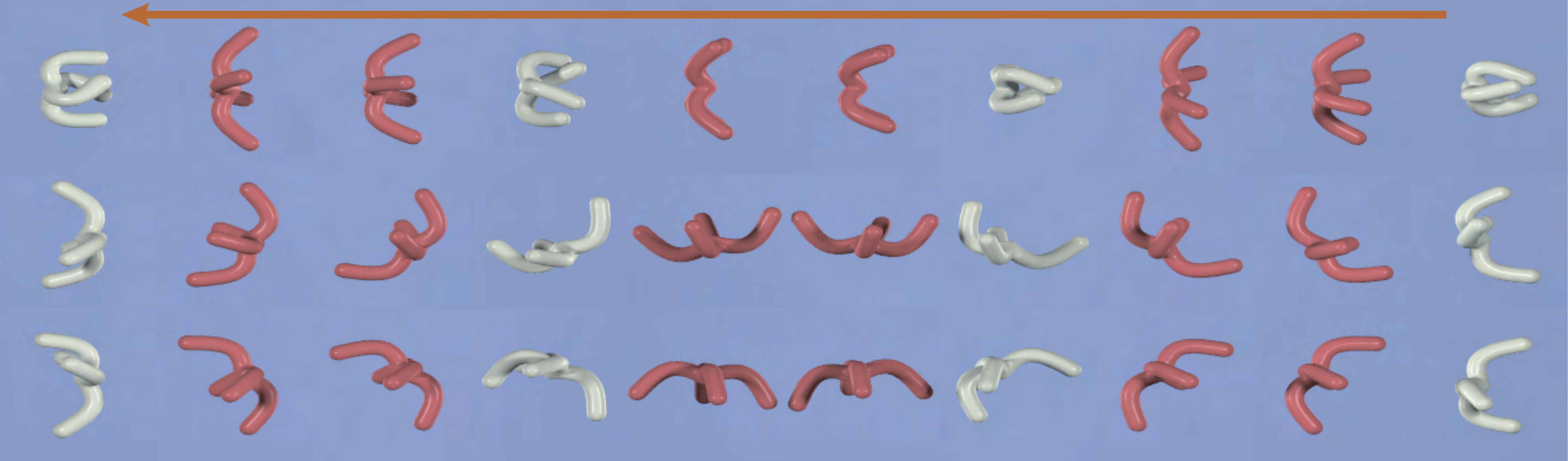}
\put(-27, 2){\small{(c)}}
\put(-27,25){\small{(b)}}
\put(-27,48){\small{(a)}}}
\end{center}
\caption{\label{fig:swimGait} To compute the navigation path for the spider, we optimize 3 trajectories: swimming forward (a), turning left (b), and turning right (c) (the timestep index increases along the arrow, and the white bodies mark the most deformed configurations). These differences in the gaits can be represented by different DMP parameters $\alpha_n$ and $\beta_n$ only.}
\end{figure}

\begin{changedBlk}
\subsection{Combining our Algorithm with Partial Keyframe Data} 
Although our main contribution is a control framework that does not require keyframes, we can easily take keyframes into consideration to provide more flexibility to a user. These keyframes can either be specified fully or partially. A full keyframe specifies a target position for each of the $V$ vertices, while a partial keyframe only specifies a target position for a subset of vertices on the deformable body. For example, in \prettyref{fig:dinoWalk} (a), we show a dinosaur walking on the ground with its head swinging periodically to the left and right. The dinosaur's head is guided by a set of $M$ partial keyframes illustrated in \prettyref{fig:dinoWalk} (b). The keyframes only specify the head and torso poses and we leave the leg poses to be determined by other objective function terms. We denote these keyframes as $u_1^\text{key},\cdots,u_M^\text{key}$ specified at timesteps $t_1,\cdots,t_M$. Note that these keyframes only specify the dinosaur's deformable poses $u$ and do not affect the global transformation $(c,w)$. The keyframe guiding is achieved using an additional objective function:
\begin{eqnarray}
\label{eq:key}
\ENGS{obj}{key}(\traj)=\COEFS{obj}{key}\sum_{i=1}^M\|I(q(u_i^\text{key})-q(u_{t_i}))\|^2/2,
\end{eqnarray}
where $I$ is an importance-weighting matrix allowing the users to specify partial keyframes. In our example, $I$ is a diagonal matrix with diagonal value $1$ around the head and torso ($788$ vertices) and $0$ elsewhere.
\begin{figure}[t]
\begin{center}
\scalebox{0.95}{
\includegraphics[width=0.49\textwidth]{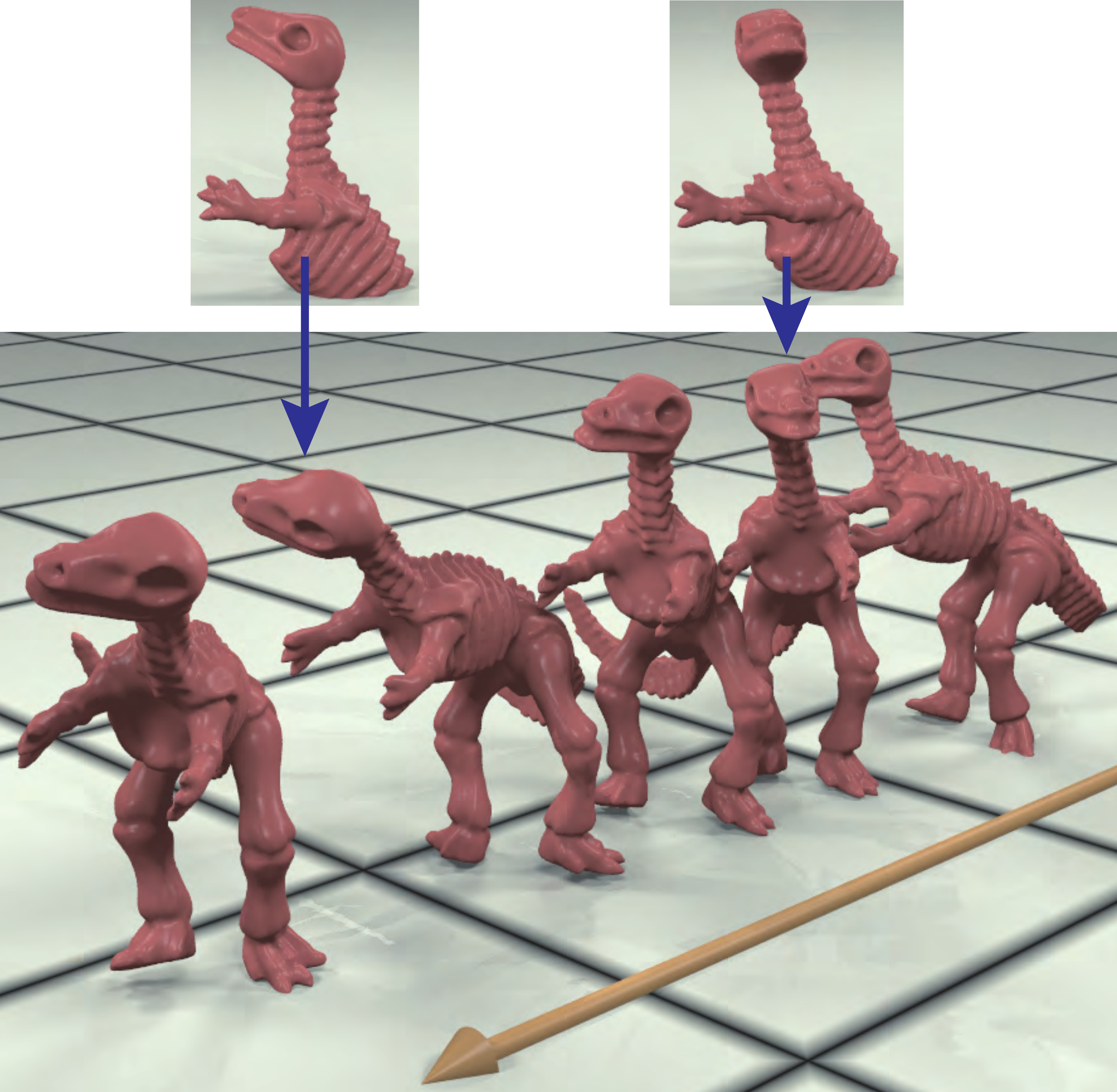}
\put(-27,15){\small{(a)}}
\put(-40,200){\small{(c)}}
\put(-150,200){\small{(b)}}}
\end{center}
\caption{\label{fig:dinoWalk} \changed{We show a walking dinosaur guided by both our high-level objectives and user-specified keyframes in (a) so that its head is looking around. And the upper-body partial keyframes are illustrated in (b,c). Each of these keyframes contains $788$ of the $1493$ vertices.}}
\end{figure}
\end{changedBlk}

\subsection{Analysis}\label{sec:analyze}
We summarize objective functions used in all benchmarks in \prettyref{table:obj} and analyze several aspects of our method.
\setlength{\columnsep}{4pt}
\begin{table}[ht]
\setlength{\tabcolsep}{4pt}
\begin{center}
\scalebox{0.8}{
\begin{tabular}{lc}
\toprule
Example          & $\ENG{obj}$   \\
\midrule
Fish Swimming    & $\ENG{obj}=\ENGS{obj}{move}+\ENGS{obj}{turn}+\ENGS{obj}{bal},$
                   $v_c=\THREE{1}{0}{0}^T,\theta=0,\pm1,d=g$   \\
Spider Swimming  & $\ENG{obj}=\ENGS{obj}{move}+\ENGS{obj}{turn}+\ENGS{obj}{bal},$ \\
                 & $v_c=\THREE{1}{0}{0}^T,\theta=0$ or
                   $v_c=\THREE{0}{0}{0}^T,\theta=\pm1,d=g$   \\
Walking          & $\ENG{obj}=\ENGS{obj}{move}+\ENGS{obj}{bal},$
                   $v_c=\THREE{\pm1}{0}{\pm1}^T,d=g$   \\
Dinosaur Walking & $\ENG{obj}=\ENGS{obj}{move}+\ENGS{obj}{bal}+\ENGS{obj}{key},$
                   $v_c=\THREE{\pm1}{0}{\pm1}^T,d=g$   \\
Jumping          & $\ENG{obj}=\ENGS{obj}{jump}+\ENGS{obj}{bal},$
                   $h=2/3/4,v_c=\THREE{2/2.5/3}{0}{0}^T,d=g$   \\
Rolling          & $\ENG{obj}=\ENGS{obj}{turn}+\ENGS{obj}{bal},$
                   $v_c=\THREE{1}{0}{0}^T,\theta=1,d=v_c\times g/\|v_c\times g\|$   \\
\midrule
                 & $\COEFS{obj}{move}=\COEFS{obj}{turn}=\COEFS{obj}{jump}=\COEFS{obj}{key}=1e^{-1}$ and $\COEFS{obj}{bal}=1e^{-2}$   \\
\bottomrule
\end{tabular}}
\end{center}
\caption{\label{table:obj} Objective function used in each benchmark.}
\end{table}

\TE{Two-Stage Algorithm:} A drawback of our method is that the optimization formulation takes a more complex form, and the resulting optimization algorithm takes longer time than \cite{Barbic:2009:DOA:1576246.1531359}. Fortunately, the DMP function returned by the optimizer can be used as a swimming controller to generate more swimming animation at realtime, as illustrated in \prettyref{fig:swim}. This makes our method much more useful than a simple keyframe interpolation. However, to generate realtime contact-rich animations, such as walking and jumping, we have to use a feedback controller instead of DMP controller. This is because the contact forces are very sensitive to the discrepancy between forward simulation model and the physics model used in spacetime optimization (model discrepancy).

\TE{Multi-Tasking:} In order to make the realtime animations directable, we need to simultaneously optimize multiple animation trajectories to allow a motion planner to pick trajectory online. However, if we sequentially run separate optimizations, the generated gaits can be quite different, e.g., the fish might swing its body with different frequencies to swim in different directions. This artifact can be mitigated if we use the same DMP parameters $\tau$ and $\mu_n$ for all the trajectories to ensure the same period of movement and phase shift, i.e., DMPs differ only in $\alpha_n$ and $\beta_n$ for different tasks. This idea has been previously used for DMP-based reinforcement learning \cite{10.3389/fncom.2013.00138}. As illustrated in \prettyref{fig:swimGait}, DMP can represent large gait differences using different $\alpha_n$ and $\beta_n$ only, while the rhythms of the movements are synchronized. We use this strategy in all the navigation examples.

\begin{figure}[t]
\begin{center}
\scalebox{0.95}{
\includegraphics[width=0.49\textwidth]{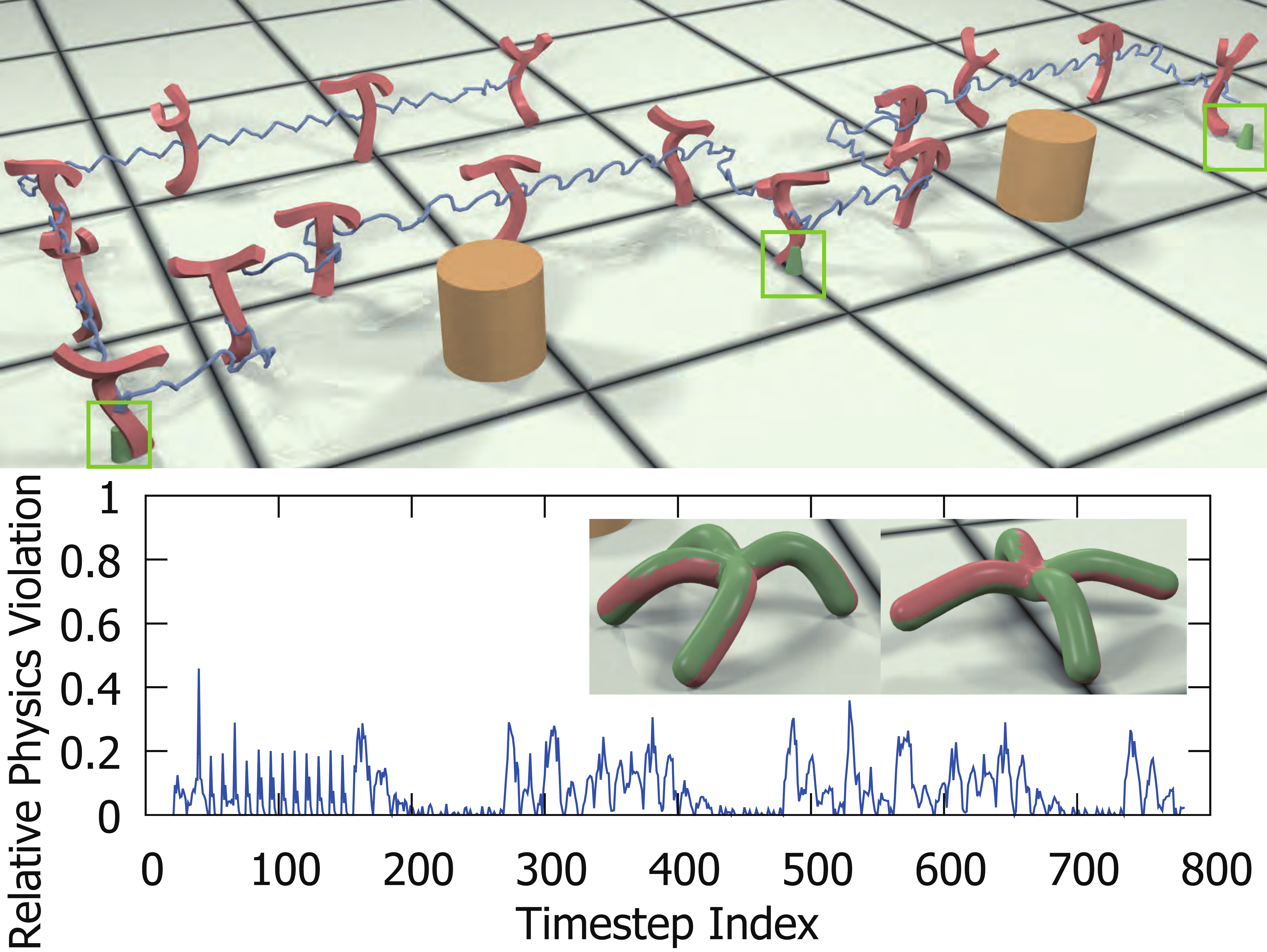}
\put(-25,100){(a)}
\put(-25,5  ){(b)}
\put(-90,55 ){(c)}}
\end{center}
\caption{\label{fig:trackT} (a): Navigating the letter T across obstacles (yellow) to reach goal positions (green). (b): A plot of the relative physical violation in (a) with respect to average FEM element size. Throughout the trajectory, the violations to EOM are very small. (c): Two most violated frames in the spider walking trajectory ($q_{i+1}^*$ drawn in green); the differences are indistinguishable.}
\end{figure}

\TE{Quality Measure:} For jumping animation, we do not require any manual bases design such as basis expansion \cite{Tan:2012:SBL:2185520.2185522}. The reason is that we formulated physics constraints as soft constraints and physics constraints are violated for small tracking errors. To measure the violation to EOM at each frame $(q_{i-1},q_i,q_{i+1})$, we first solve \prettyref{eq:eom} using $q_{i-1},q_i$ to find a physically correct $q_{i+1}^*$. Next, we measure the relative error against the average element size in Euclidean space using:
\begin{eqnarray*}
\sqrt{\|\bar{q}(q_{i+1})-\bar{q}(q_{i+1}^*)\|^2/V}/l.
\end{eqnarray*}
According to the plot in \prettyref{fig:trackT}, the physics violation over the whole trajectory is always less then half of average element size and is neglectible. The physics violation data for other examples can be found in \refSuppPhysVio. However, manual bases design can sometimes be needed. For example, very different rolling gaits are generated in \prettyref{fig:roll}, by restricting the bases to the 2D plane. 

\begin{changedBlk}
\TE{Effect of Different Parameters:} Instead of using keyframes, the result of our algorithm depends on two sets of parameters. A first set of parameters are listed in \prettyref{table:param}. These parameters are considered internal and not exposed to users. The second set of parameters listed in \prettyref{table:obj} are exposed to users. These parameters have clear meanings such as walking, swimming, or rolling speed. In \prettyref{fig:paramSens}, we highlight the effectiveness of performing animation control using the parameters listed in \prettyref{table:obj}. We generated 9 walking/swimming trajectories using different target moving speed $v_c$. Since we model $E_{obj}$ as a soft penalty, the desired speed cannot be achieved exactly. However, according to \prettyref{fig:paramSens}, the discrepancy between actual and desired moving speeds are very small. Therefore, we expose more parameters to the users compared with keyframe-based methods \cite{Barbic:2009:DOA:1576246.1531359,Schulz:2014:ADO:2601097.2601156}, these parameters have intuitive meanings and are helpful for animation control.

We also noticed two cases from \prettyref{fig:paramSens} (green circles) where the discrepancy between desired and actual moving speed are relatively large. If the desired speed is too small, then our optimizer considers $\ENG{obj}$ as unimportant and it is given lower importance in order to reduce the residue in other objective terms. If the desired speed is too large, it can result in self-collisions or the optimizer falls into a bad local minima, as shown in \prettyref{fig:failure}.
\begin{figure}[t]
\begin{center}
\scalebox{0.95}{
\includegraphics[width=0.49\textwidth]{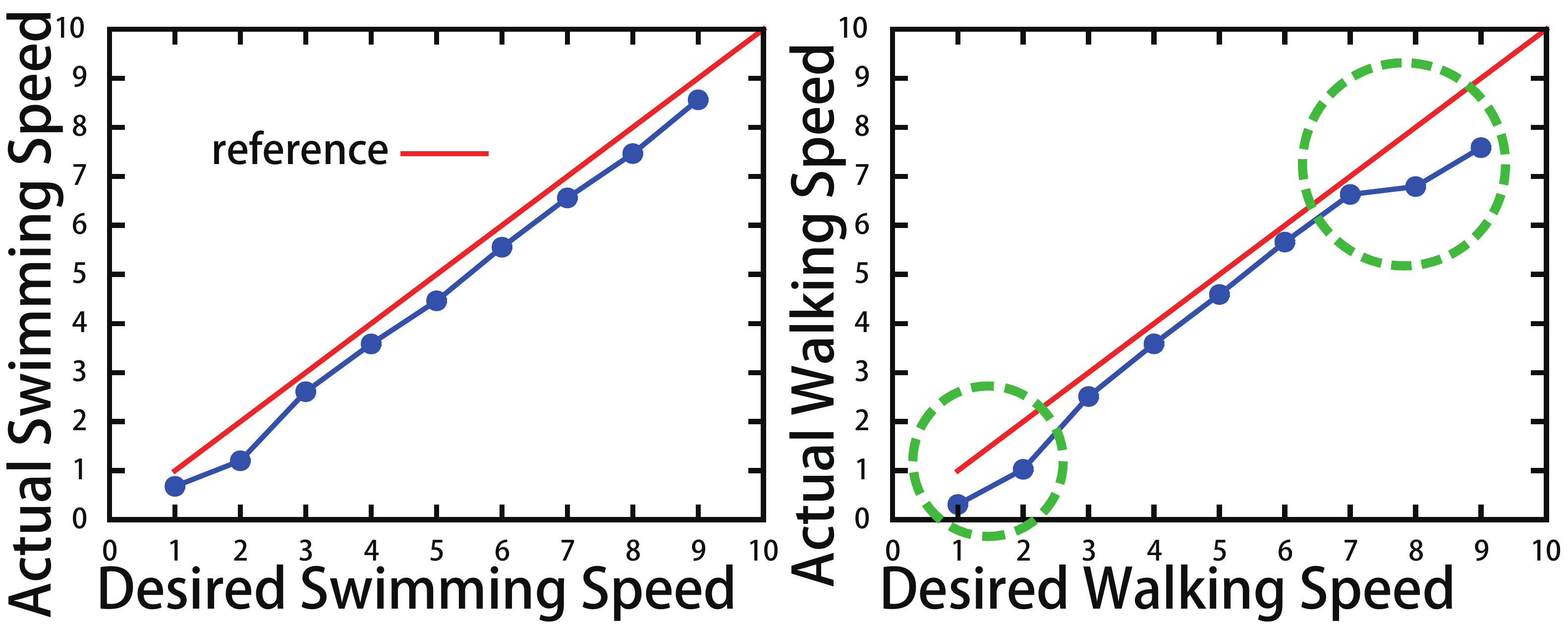}
\put(-150,40){(a)}
\put(-25 ,40){(b)}}
\end{center}
\caption{\label{fig:paramSens} \changed{We generate $9$ spider walking trajectories (a) and $9$ spider swimming trajectories (b), using different target moving speed $v_c$. We plot the actual walking/swimming speed compared with the desired speed. These actual values are very close to desired values. However, there are still cases where the discrepancy between actual and desired values are large (green circles).}}
\end{figure}

\TE{External vs. Internal Control Forces:} Theoretically, our algorithm only uses internal forces to control the deformable body because our control force $\fint{}{}$ is dot-producted only with $u$, instead of the entire $q$ in \prettyref{eq:eom}. Therefore, it does not change the global translation $c$ and rotation $w$. However, since we formulate the physical correctness as a penalty term, $\ENG{eom}$, rather than a hard constraint, there is some residual $\ENG{eom}\neq 0$ at the local minima. This residual can be interpreted as a violation of the physical correctness, or as a ghost external force. If we write $\ENG{eom}(q_{i-1},q_i,q_{i+1})=\|\fext{i}{ghost}^T\bar{q}\|$, we are actually controlling the deformable body using both internal force $\fint{i}{}$ and an additional ghost external force $\fext{i}{ghost}$, but our objective function is designed to guide the optimizer to search for a solution with minimal ghost external force magnitude. 

Using soft penalty instead of hard constraints also allows us to generate realtime deformable body animations by tracking an optimized animation. For example, having the letter-T balanced on a single contact point in \prettyref{fig:trackT} is very challenging, which usually requires control over long horizons. However, with soft penalty, we can track the animation by control over only one timestep using \prettyref{eq:objTrack} as the objective function.

\TE{Robustness of Reduced Model Construction:} A critical step in our method is the construction of the reduced model. Although this procedure takes several parameters, prior work \cite{An:2008:OCE:1457515.1409118,Pan:2015:SDS:2816795.2818090} have resulted in robust algorithms that perform consistently well on a large dataset using same parameters. Therefore, we consider the construction of reduced model fully automatic. For example, to select the set of surface patches in \prettyref{eq:cubature} and to construct the transformation function $\bar{q}(q)$, we use cubature optimization. This procedure requires a training dataset. Our dataset is constructed by sampling 1000 Gauss-distributed deformable body poses as suggested in \cite{An:2008:OCE:1457515.1409118}. In order to solve the dictionary learning for the set of cubatures ($\mathcal{T}$ in \prettyref{eq:cubature}), we use L0-optimization proposed in \cite{Pan:2015:SDS:2816795.2818090} which automatically determine the required number of cubatures. We follow \cite{Pan:2015:SDS:2816795.2818090} in all other parameter settings and have never observed any failure cases.

\TE{Other Reduced Models:} Although we choose \cite{Pan:2015:SDS:2816795.2818090}, our method can also work with other reduced models. This can be done by modifying the transformation function $\bar{q}(q)$ and the kinetic energy $P$ in \prettyref{eq:eom}. In \prettyref{appen:otherRM}, we analyze the case with two kinds of different but widely used reduced models: LMA \cite{Pentland:1989:GVM:74334.74355} and reduced StVK \cite{Barbic:2005:RSI:1186822.1073300}. And two examples are illustrated in \prettyref{fig:STVKLetterT} (a,b). 

A drawback of these alternative models is that they require a higher-dimensional configuration space to achieve similar results as \cite{Pan:2015:SDS:2816795.2818090}. In our experiments, we use $|u|=65$ and each optimization becomes 3-5 times slower according to \prettyref{table:perf}. However, from the plots of DMP control force magnitudes, \prettyref{fig:STVKLetterT} (c,d), we notice that the optimal $\fint{}{}$ is actually very sparse. In other words, much computations are wasted on looking for small, unimportant control forces. Such analysis suggests that \cite{Pan:2015:SDS:2816795.2818090} is a better choice.

Although our formulation can also work with fullspace deformable models by replacing $\bar{q}$ with identity function, this approach can be computationally very expensive. As reported in \cite{Pan:2015:SDS:2816795.2818090}, using a reduced model accelerates the evaluation of $\bar{q}(q)$ by two orders of magnitude. In our experiments, cubature accelerates the evaluation of fluid drag forces by at least an order of magnitude using \prettyref{eq:cubature}. Since function evaluation is the major bottleneck of spacetime optimization, we expect it will take weeks or even months to finish an optimization using fullspace models. As a result, it is important to use reduced deformable models for efficiency reasons.
\begin{figure}[t]
\begin{center}
\scalebox{0.95}{
\includegraphics[width=0.49\textwidth]{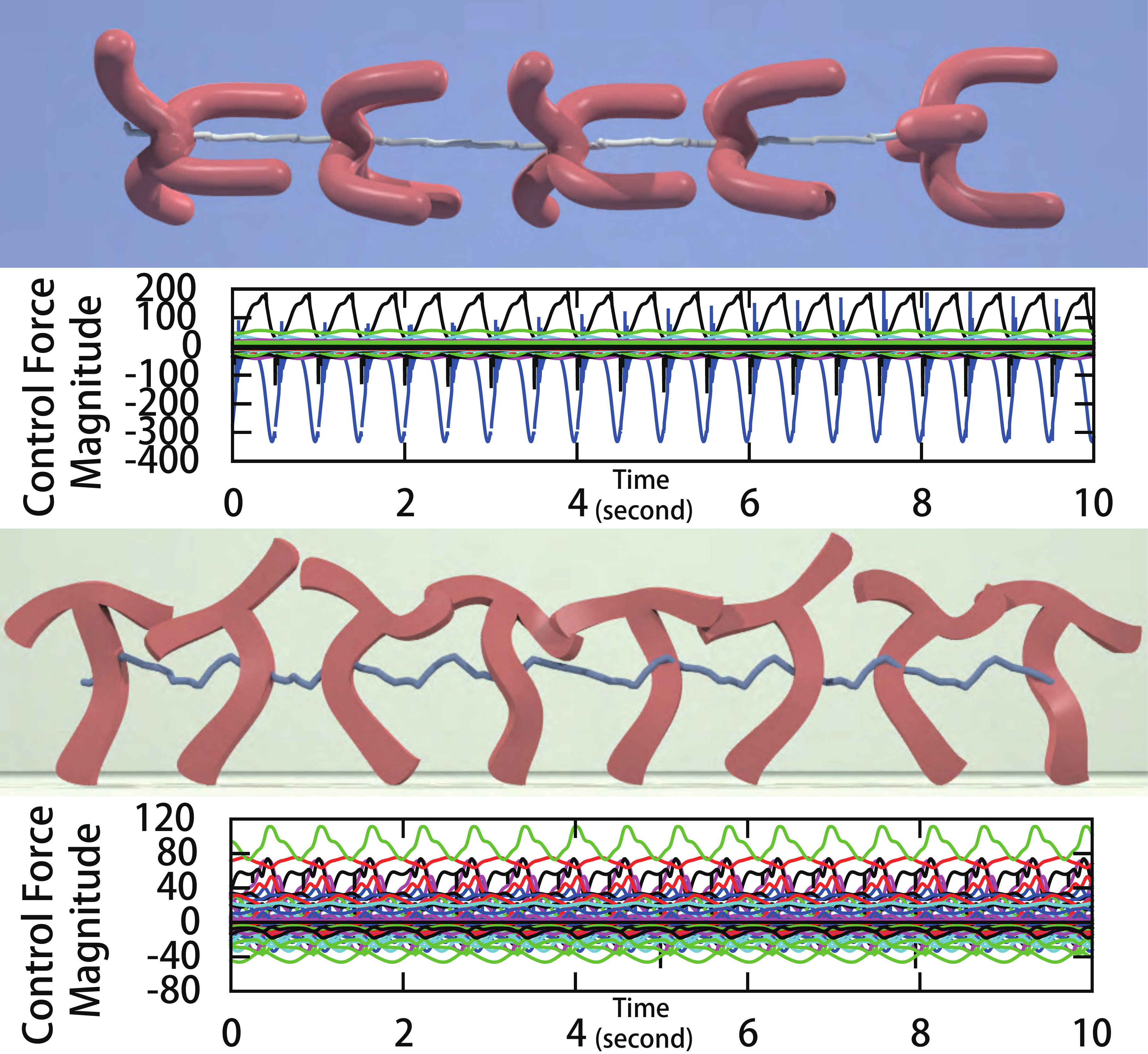}
\put(-15,220){(a)}
\put(-15,105){(b)}
\put(-8,145){(c)}
\put(-8,30){(d)}}
\end{center}
\caption{\label{fig:STVKLetterT} \changed{We computed the animations corresponding to spider swimming (a) and letter T walking (b) using reduced StVK \cite{Barbic:2005:RSI:1186822.1073300} as the underlying deformable model. In addition, we plot the control force magnitude for the spider in (c) and for the letter T in (d). We use different curves to plot each $\fint{}{j}$. Since $|u|=65$, we have $65$ curves in each plot. However, most of these curves are centered around zero axis. This means that the magnitude of control forces for most $j$ are very small and $\fint{}{}$ is quite sparse.}}
\end{figure}
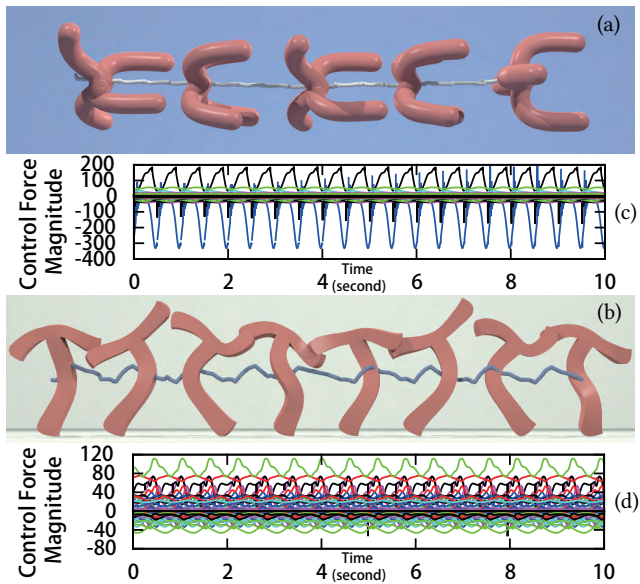
\end{changedBlk}

\section{Limitations and Future Work}\label{sec:conclusion}
We present a method to automatically generate active animations of reduced deformable bodies, where the user provides a high-level objective and the animation is generated automatically using spacetime optimization. We take into account physics constraints, environmental forces in terms of CIO and fluid drag models, and DMP-based controller parametrization, so that the local minima of our objective function corresponds to a plausible animation. By evaluating objective functions and function gradients in a subspace, the optimization can be accomplished within several hours on a single desktop machine. Although optimization is offline, the results can be used to generate animations at realtime rates. For swimming animations, the optimized DMPs can be used as a controller for forward simulation. Unfortunately, DMP cannot be used as controllers for contact-rich animations. Since DMP is not a feedback controller, model discrepancy can quickly accumulate, leading to failures such as falling. In these cases, DMP is just used as a periodic and smoothness prior. 

Our approach has some limitations. \changed{First, our method inherits all the limitations of the underlying reduced model. For example, current reduced model~\cite{Pan:2015:SDS:2816795.2818090} cannot work with user specified skeletons. Working with skeletons is a desirable feature in terms of modelling some animal-like deformable bodies, such as the fish, where deformable tissues are covering skeletal bones. In addition, although our method requires no keyframes or user designs, we still ask the users to choose the form of $\ENG{obj}$ and their parameters in \prettyref{sec:results}. And without keyframes, the animations may not exhibit the same level of naturalness as some prior keyframe-based methods~\cite{Barbic:2009:DOA:1576246.1531359}.} Moreover, our optimizer may get stuck in a bad local minima due to insufficient DOFs of the reduced configuration space, a sub-optimal bases set, or an inappropriate settings of the weights. \changed{Furthermore, the inherent limitations of CIO term \cite{Mordatch:2013:AHL:2508363.2508365} for contact modeling and the fluid drag model can also affect our results. For example, we cannot have a deformable body bouncing off the ground since the CIO term only models inelastic contacts. CIO also allows inexact contacts to occur anywhere in the air, not only on the ground.} Finally, like all the optimization-based motion planners, the performance of our method is still governed by a large set of parameters. Some parameters, such as the number of DMPs $(\mathcal{N})$, are determined empirically. We have not evaluated the sensitivity of our method with respect to these parameters.

There are avenues for future work. First, incorporating some body-specific priors can be helpful in several ways. For example, for many muscle-driven deformable bodies, the user might want to parameterize the controller using muscle-tendon units~\cite{Wang:2012:OLC:2185520.2185521} to generate more life-like animations. Another part that may benefit from user interactions is the identification of deformation bases in \prettyref{fig:reduced}. Currently, we identify these components using standard techniques~\cite{Pan:2015:SDS:2816795.2818090}  that are designed for visual simulation. However, it is not known if a base set for plausible visual simulation is suitable for character locomotion. \changed{It is also attractive to consider the optimization method as a general feedback controller, instead of an open-loop controller, for reduced deformable models using reinforcement learning~\cite{Peng:2017:DDL:3072959.3073602}. Finally, developing control methods for two-way couple deformable body and articulated body will provide more flexibility to users. A starting point can be \cite{Xu:2016:PSD:2897824.2925916}.}
\bibliographystyle{ACM-Reference-Format}
\bibliography{template}
\appendix
\section{Optimization Algorithm}\label{appen:OA}
In this section we summarize our main \prettyref{alg:opt}. The skeleton of this algorithm is an LM optimizer. We refer readers to \cite{lourakis2005brief} for a brief introduction of this simple method. In our experiments, LM performs much better than LBFGS for two reasons. First, LM uses $J^TJ$ approximation of the hessian which usually leads to a better step size estimation. Second, although LM needs to search for the so-called damping coefficient as in the process of the line-search scheme, it requires only function value evaluations which are much faster than gradient evaluations used by the line-search scheme of LBFGS. 

However, a minor problem with using an LM algorithm is that it assumes the objective function is a sum of squares, which is violated by $\ENG{shuffle}$. In \prettyref{appen:LMM}, we show that a marginal modification to \cite{lourakis2005brief} will allow us to handle $\ENG{shuffle}$.

After every successful LM iteration, we update $\fext{}{}$; we update $\DMPW{}$ every 10 iterations. Since each update leads to an energy value decrease, the algorithm is guaranteed to converge eventually. Finally, we also update the active vertices contributing to $\ENG{env}$ and $\ENG{shuffle}$ in \prettyref{ln:cio}. A minor bottleneck in this algorithm is the sparse linear system in \prettyref{ln:linsol}, but we notice that the hessian of $E(\traj)$ is a block-tridiagonal matrix that can be inverted with linear time complexity.

\subsection{LM Modification for \texorpdfstring{$\ENG{shuffle}$}{Lg}}\label{appen:LMM}
Conventional LM assumes that the objective function takes the form of $E(\traj)=f(\traj)^Tf(\traj)/2$ where $f$ is a vector of nonlinear terms. This is not the case with $\ENG{shuffle}$. However, among the many implementations of LM, the one documented in \cite{lourakis2005brief} is not limited to the above form. If we have a certain approximation of the hessian of $E(\traj)$ denoted as $H$ (not necessarily in the form of $J^TJ$) and use damping coefficient $d$, then the update to $\traj$ in our main algorithm is computed as:
\begin{eqnarray*}
\Delta\traj=-(H+d\E{I})^{-1}\FPP{E(\traj)}{\traj},
\end{eqnarray*}
and the decrease in $E(\traj)$ after applying the update is:
\begin{eqnarray*}
\Delta\traj^T\FPP{E(\traj)}{\traj}+\frac{1}{2}\Delta\traj^TH\Delta\traj
=\frac{1}{2}\Delta\traj^T(\FPP{E(\traj)}{\traj}-d\Delta\traj),
\end{eqnarray*}
which is exactly the expected function value decrease estimation used in \cite{lourakis2005brief}. This allows us to use a non-$J^TJ$ form of approximate hessian for $\ENG{shuffle}$. In our case, we approximate the term $\ENG{shuffle}$ by first order expansion for both $v^j(q_i)$ and $v^j(q_{i-1})$, giving:
\begin{eqnarray*}
&&\ENG{shuffle}(q_i,q_{i-1}) \\
&=&\sum_{j=1}^V\|(v^j(q_i+\Delta q_i)-v^j(q_{i-1}+\Delta q_{i-1}))_\parallel\|^2 \\
&&\E{exp}(-\gamma\E{dist}(v^j(q_i+\Delta q_i))) \\
&\approx&\sum_{j=1}^V\|(v^j(q_i)+\FPP{v^j}{q_i}\Delta q_i-v^j(q_{i-1})-\FPP{v^j}{q_{i-1}}\Delta q_{i-1})_\parallel\|^2 \\
&&\E{exp}(-\gamma\E{dist}(v^j(q_i)+\FPP{v^j(q_i)}{q_i}\Delta q_i)),
\end{eqnarray*}
and use the hessian of the last equation above.
Conventional $J^TJ$ approximation of the hessian is used for all other terms.

\subsection{Eliminating Internal Force Terms}\label{appen:EIFT}
In our objective function, $\fint{i}{}$ is a quadratic function in $\ENG{eom}$, in the Tikhonov regularization $\COEF{reg}\|\fint{i}{}\|^2/2$, and in the DMP regularization $\ENG{dmp}$. We analytically solve and eliminate $\fint{i}{}$ from the optimization as follows:
\begin{small}
\begin{eqnarray*}
&&\fmin{\fint{i}{}}\ENG{eom}(q_{i-1},q_i,q_{i+1})+\COEF{reg}\|\fint{i}{}\|^2/2+\COEF{dmp}{\ENG{dmp}}(\fint{i}{})  \\
&=&\fmin{\fint{i}{}}\frac{1}{2}\left[\|EOM_i-\fint{i}{}\|^2+\COEF{reg}\|\fint{i}{}\|^2+\COEF{dmp}\|\fint{i}{}-DMP_i\|^2\right]  \\
&=&\frac{1}{2}\TWO{EOM_i^T}{DMP_i^T}\MASS{reg}\TWOC{EOM_i}{DMP_i}  \\
&&EOM_i\triangleq\THREE{\E{I}}{0}{0}\left[\FPP{\bar{q}}{q}^T(q_{i+1})MA(q_{i+1})+\FPP{\left[P-\fext{i}{}^T\bar{q}\right]}{q}(q_{i+1})\right] \\
&&DMP_i\triangleq\FOURC{DMP_{p/np}(i\Delta t,\DMPW{1})}{DMP_{p/np}(i\Delta t,\DMPW{2})}{\vdots}{DMP_{p/np}(i\Delta t,\DMPW{|\fint{i}{}|})}  \\
&&\MASS{reg}\triangleq\frac{1}{1+\COEF{reg}+\COEF{dmp}}\MTT{\COEF{reg}+\COEF{dmp}}{-\COEF{dmp}}{-\COEF{dmp}}{\COEF{reg}\COEF{dmp}+\COEF{dmp}}\otimes\E{I}.
\end{eqnarray*}
\end{small}
\section{\texorpdfstring{$v(q)$}{Lg} and its Derivatives}\label{appen:VAID}
In this section, we briefly summarize the transformation function from the reduced representation $q$ to a vertex $v$'s Euclidean coordinates as defined in \cite{Pan:2015:SDS:2816795.2818090}.  We then provide some guidance on the computation of $\FPP{v(q)}{q}$ and $\FPPT{v(q)}{q}$. Since we use the LM algorithm for space-time optimization, most objective terms only require a first order derivative $\FPP{v(q)}{q}$. An exception is the physics violation term $\ENG{eom}$, which requires second order derivatives.

\subsection{Transformation Function}\label{appen:TF}
We assume that a deformable body is discretized using $V$ vertices and $P$ elements. For each element $j$, its deformation gradient is denoted as $F^j$. Vertices' Euclidean coordinates $\bar{q}$ can be reconstructed from $F^j$ through Poisson reconstruction. We abbreviate this linear operator as:
\begin{eqnarray*}
\bar{q}=\Delta^{-1}\THREE{{F^1}^T}{\cdots}{{F^P}^T}^T.
\end{eqnarray*}
It is well-known that $F^j$ above has a polar decomposition $F^j=R^jS^j$ where $R^j$ is a rotation and $S^j$ is an anisotropic scaling. We can then use the Rodrigues formula on the rotation part to get $R^j=\E{exp}(\tau^j)$, where $\tau^j$ is the rotation vector of element $j$. The rotation-strain (RS) space is defined by the space spanned by all possible $\tau^j$ and $S^j-\E{I}$:
\begin{eqnarray*}
\E{span}(RS)=\{(\tau^1,\cdots,\tau^P,S^1-\E{I},\cdots,S^P-\E{I})|S^j \text{is SPD}\}.
\end{eqnarray*}
RS representation and Euclidean coordinates are equivalent. However, using RS representation is advantageous because the most visually salient deformations lie in a low-dimensional linear subspace of $\E{span}(RS)$. Therefore, we can use conventional linear dimensionality reduction techniques such as linear modal analysis \cite{Pentland:1989:GVM:74334.74355} in RS space to arrive at the following low-rank approximation:
\begin{eqnarray*}
(\tau^1,\cdots,\tau^P,S^1-\E{I},\cdots,S^P-\E{I})\approx Bu,
\end{eqnarray*}
where $B$ is a set of bases in the RS subspace. The transformation function from $u$ to $\bar{q}$ can then be defined by combining the above equations:
\begin{small}
\begin{eqnarray*}
&&\bar{q}(q)=\Delta^{-1}[\E{exp}](Bu)[\E{S}](Bu) \\
&&\left[\E{exp}\right](Bu)\triangleq\MDD{\E{exp}(\tau^1)}{\ddots}{\E{exp}(\tau^P)}
\quad\quad
\left[\E{S}\right](Bu)\triangleq\THREEC{S^1}{\vdots}{S^P}.
\end{eqnarray*}
\end{small}
However, this only encodes deformations in the local frame of reference. To allow arbitrary movement in the global frame of reference, we can superimpose a global translation $c$ and rotation $w$, giving:
\begin{small}
\begin{eqnarray*}
\bar{q}(q)=
\MDD{\E{exp}(w)}{\ddots}{\E{exp}(w)}
\Delta^{-1}[\E{exp}](Bu)[\E{S}](Bu)+
\THREEC{c}{\vdots}{c}.
\end{eqnarray*}
\end{small}

\subsection{Kinetic Cubature Acceleration}\label{appen:KCA}
Merely having the low-rank approximation does not accelerate computation. The evaluation of $\bar{q}$ is still computationally costly because it requires a summation over all the $P$ elements. This procedure can be accelerated using cubature approximation by assuming the following approximation:
\begin{eqnarray*}
&&\Delta^{-1}[\E{exp}](Bu)[\E{S}](Bu) \\
&=&\sum_{j}\Delta_j^{-1}\E{exp}(w^j)S^j
\approx\sum_{j\in\mathcal{T}}\COEF{RS}^j\Delta_j^{-1}\E{exp}(w^j)S^j,
\end{eqnarray*}
where $\Delta_j^{-1}$ is the block of $\Delta^{-1}$ corresponding to the $j$th element. The weighting $\COEF{RS}^j$ and the set of cubature elements $\mathcal{T}$ are precomputed using L0-optimization.

\subsection{Derivatives}\label{appen:DERIV}
To derive $\FPP{v(q)}{q}$ and $\FPPT{v(q)}{q}$, we notice that they can always be written as a long chain of matrix productions, where each matrix is either a constant, a linear function of $q$, or a linear function of $\E{exp}(w)$, $\FPP{\E{exp}(w)}{w}$, and $\FPPT{\E{exp}(w)}{w}$. Equations of this form can be accelerated using fast sandwich transform (FST) \cite{Kim:2011:PCS:2019406.2019415}, i.e., by precomputing high-order tensors and contracting them with $q,\E{exp}(w)$, $\FPP{\E{exp}(w)}{w}$, and $\FPPT{\E{exp}(w)}{w}$ at runtime. The remaining problem is to derive $\FPP{\E{exp}(w)}{w}$ and $\FPPT{\E{exp}(w)}{w}$. A closed-form of $\FPP{\E{exp}(w)}{w}$ can be found in \cite{2015-03-MIV2014_Gallego}. We now derive $\FPPT{\E{exp}(w)}{w}$ below using their notations:
\begin{eqnarray*}
&&\E{R}\triangleq\E{exp}(w)\quad
\FPP{\E{R}}{w_i}=\CROSS{\E{v}_i}\E{R} \\
&&\FPPM{\E{R}}{w_i}{w_j}=\CROSS{\FPP{\E{v}_i}{w_j}}\E{R}+\CROSS{\E{v}_i}\FPP{\E{R}}{w_j} \\
&&\E{v}_i\triangleq\frac{w_iw+\CROSS{w}(\E{I}-\E{R})e_i}{\|w\|^2} \\
&=&\bar{w}_i\bar{w}-\frac{\E{sin}(\theta)\CROSS{\bar{w}}^2+(1-\E{cos}(\theta))\CROSS{\bar{w}}^3}{\theta}e_i \\
&=&\frac{\theta-\E{sin}(\theta)}{\theta}\bar{w}_i\bar{w}+\frac{1-\E{cos}(\theta)}{\theta}\CROSS{\bar{w}}e_i+\frac{\E{sin}(\theta)}{\theta}e_i,
\end{eqnarray*}
where we used the identity $\bar{w}=w/\|w\|$, $\theta=\|w\|$, $\CROSS{\bar{w}}^2=(\bar{w}\bar{w}^T-\E{I})$, and $\bar{w}^T\CROSS{\bar{w}}=\E{0}$. Finally, the $\FPP{\E{v}_i}{w_j}$ above has the following form:
\begin{eqnarray*}
&&\FPP{\E{v}_i}{w_j}=\frac{\theta-\E{sin}(\theta)}{\theta^2}
(e_j\bar{w}_i+\bar{w}\delta_{ij}-2\bar{w}\bar{w}_i\bar{w}_j)+   \\
&&\frac{\E{sin}(\theta)-\theta\E{cos}(\theta)}{\theta^2}\bar{w}\bar{w}_i\bar{w}_j+
\frac{1-\E{cos}(\theta)}{\theta^2}\CROSS{e_j-\bar{w}_j\bar{w}}e_i+  \\
&&\frac{\theta\E{sin}(\theta)+\E{cos}(\theta)-1}{\theta^2}\CROSS{\bar{w}}e_i\bar{w}_j-
\frac{\E{sin}(\theta)-\theta\E{cos}(\theta)}{\theta^2}e_i\bar{w}_j.
\end{eqnarray*}
\begin{changedBlk}
\section{Other Reduced Models}\label{appen:otherRM}
In this section, we analyze the cases where rotation-strain coordinates \cite{Pan:2015:SDS:2816795.2818090} is replaced with either LMA \cite{Pentland:1989:GVM:74334.74355} or reduced StVK \cite{Barbic:2005:RSI:1186822.1073300}. As mentioned in \prettyref{sec:analyze}, we need to modify both $\bar{q}(q)$ and $P$ in \prettyref{eq:eom}. In both cases, $\bar{q}(q)$ takes the following simple form:
\begin{small}
\begin{eqnarray*}
\bar{q}(q)=\MDD{\E{exp}(w)}{\ddots}{\E{exp}(w)}(Uu)+\THREEC{c}{\vdots}{c},
\end{eqnarray*}
\end{small}
where $U$ is a set of linear deformation bases. Due to the lack of rotation-strain transformation, we have to introduce more columns to $U$ than $B$ in order to represent nonlinear deformations. If LMA is used, $P$ takes the same quadratic form as that for rotation-strain coordinates, $P(q)=u^T\mathcal{K}u/2$. If reduced StVK is used, $P(u)$ is a quartic function in $u$ whose polynomial coefficients can be precomputed.
\end{changedBlk}
\section{Physics Violation}\label{appen:PV}
We provide the physics violation data for all the examples in \prettyref{table:physVio}. We use the same criterion:
\begin{eqnarray*}
\sqrt{\|\bar{q}(q_{i+1})-\bar{q}(q_{i+1}^*)\|^2/V}/l.
\end{eqnarray*}
Compared with the average FEM element size $l$, the error due to physics violation is very small. The parameters for all the volumetric deformable models are provided in \prettyref{table:model}.
\begin{table}[ht]
\setlength{\tabcolsep}{10pt}
\begin{center}
\scalebox{0.95}{
\begin{tabular}{lll}
\hline
Model & $(X\times Y\times Z)$ & l   \\
\hline
Fish & $2\times1.2\times0.2$ & $0.071$    \\
Spider & $1.65\times0.51\times1.65$ & $0.079$   \\
Letter T & $1.00\times1.24\times0.1$ & $0.037$    \\
Dragon & $1.52\times0.71\times1.03$ & $0.075$   \\
Beam & $0.2\times2\times0.2$ & $0.14$    \\
Cross & $1\times1\times0.1$ & $0.042$   \\
\hline
\end{tabular}}
\end{center}
\caption{\label{table:model} Model parameters used in our experiment. From left to right, model name, bounding box size $(X\times Y\times Z)$, and average FEM element size l.}
\end{table}

\begin{table}[ht]
\setlength{\tabcolsep}{2pt}
\begin{center}
\scalebox{0.95}{
\begin{tabular}{ll}
\hline
Fish Swimming & Spider Swimming \\
\hline
\includegraphics[trim={0 2cm 0 2cm},clip,width=0.25\textwidth]{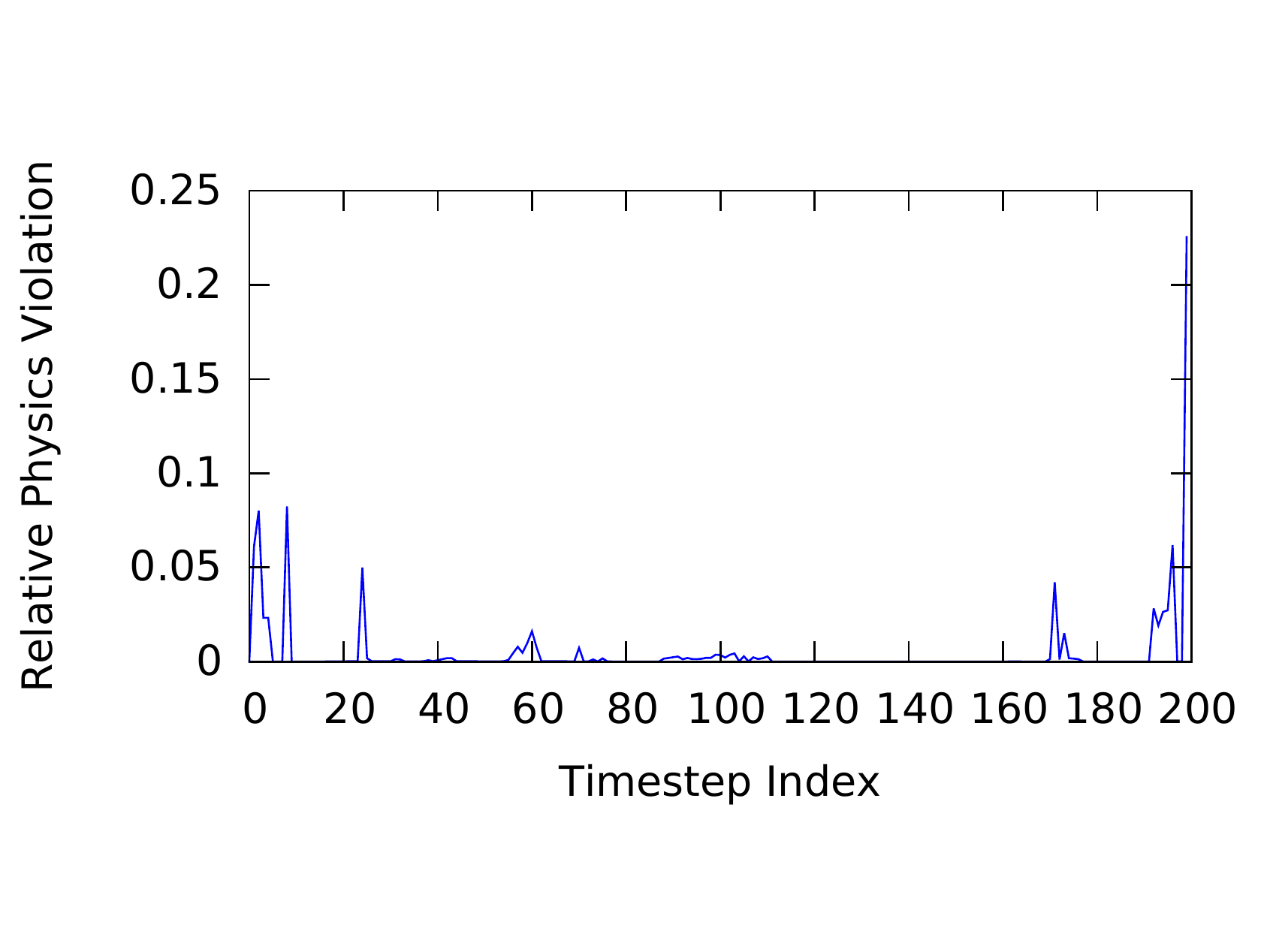} &
\includegraphics[trim={0 2cm 0 2cm},clip,width=0.25\textwidth]{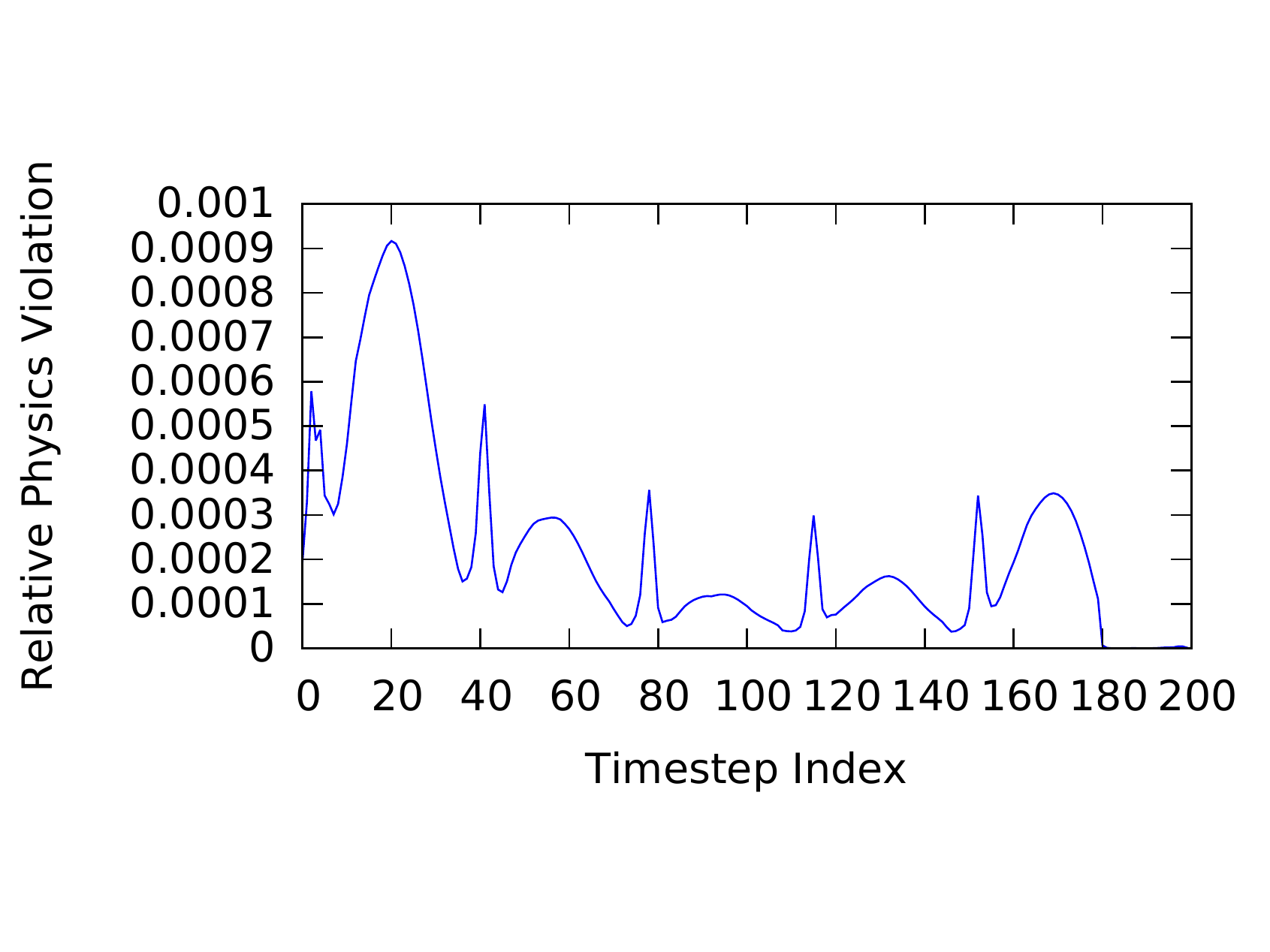} \\
\hline
Spider Walking & Letter T Walking \\
\hline
\includegraphics[trim={0 2cm 0 2cm},clip,width=0.25\textwidth]{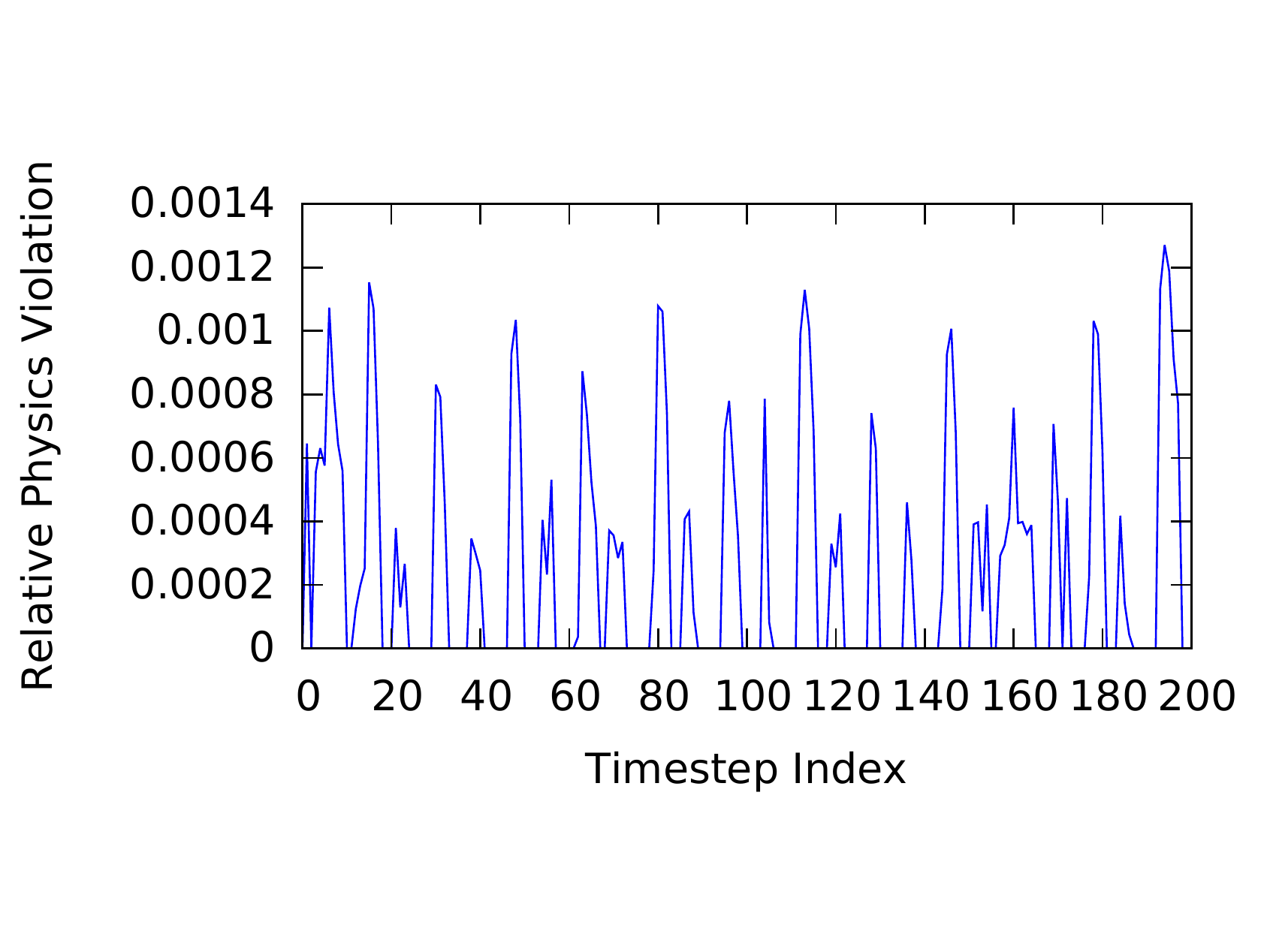} &
\includegraphics[trim={0 2cm 0 2cm},clip,width=0.25\textwidth]{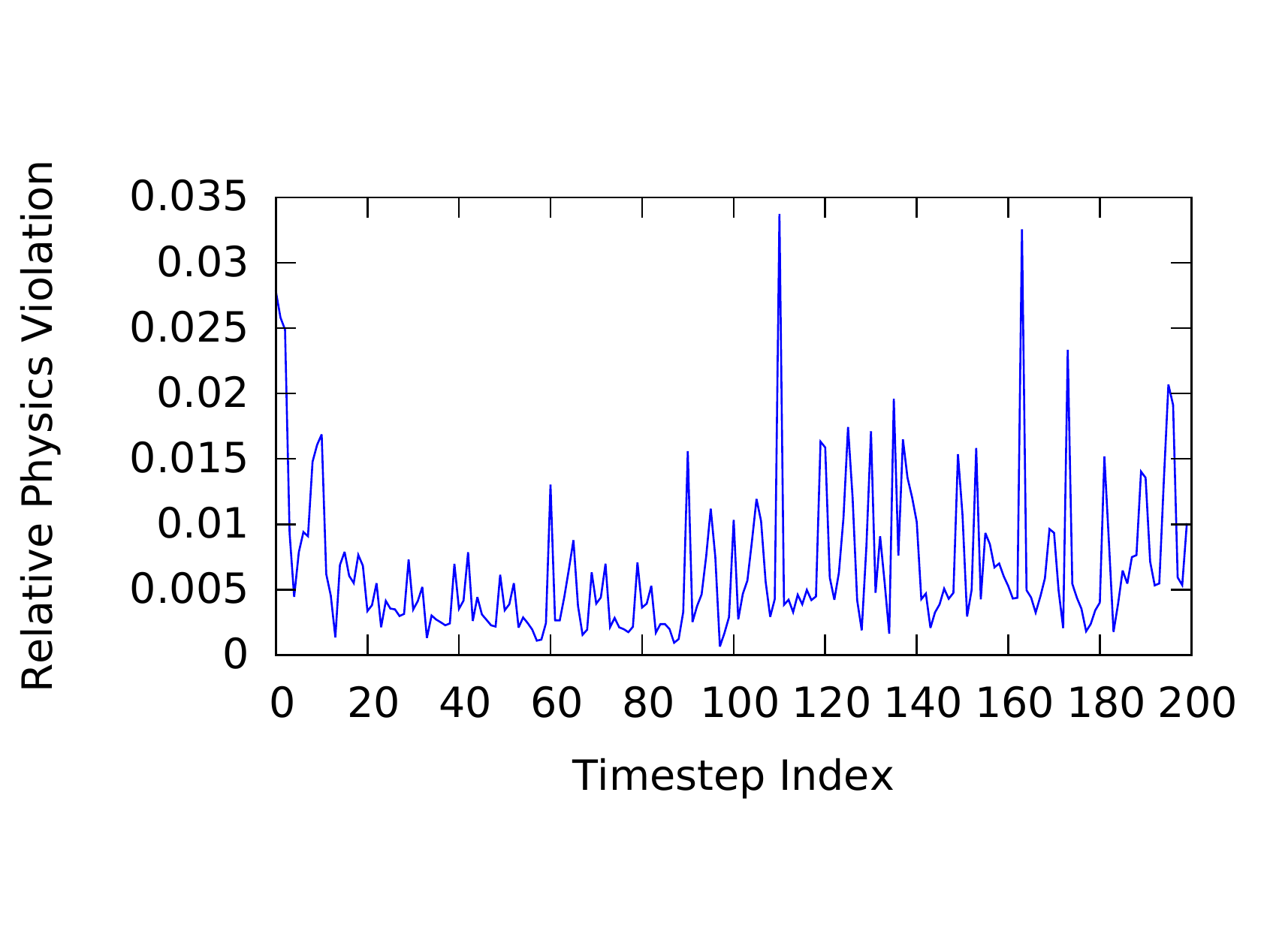} \\
\hline
Dragon Walking & Beam Jumping \\
\hline
\includegraphics[trim={0 2cm 0 2cm},clip,width=0.25\textwidth]{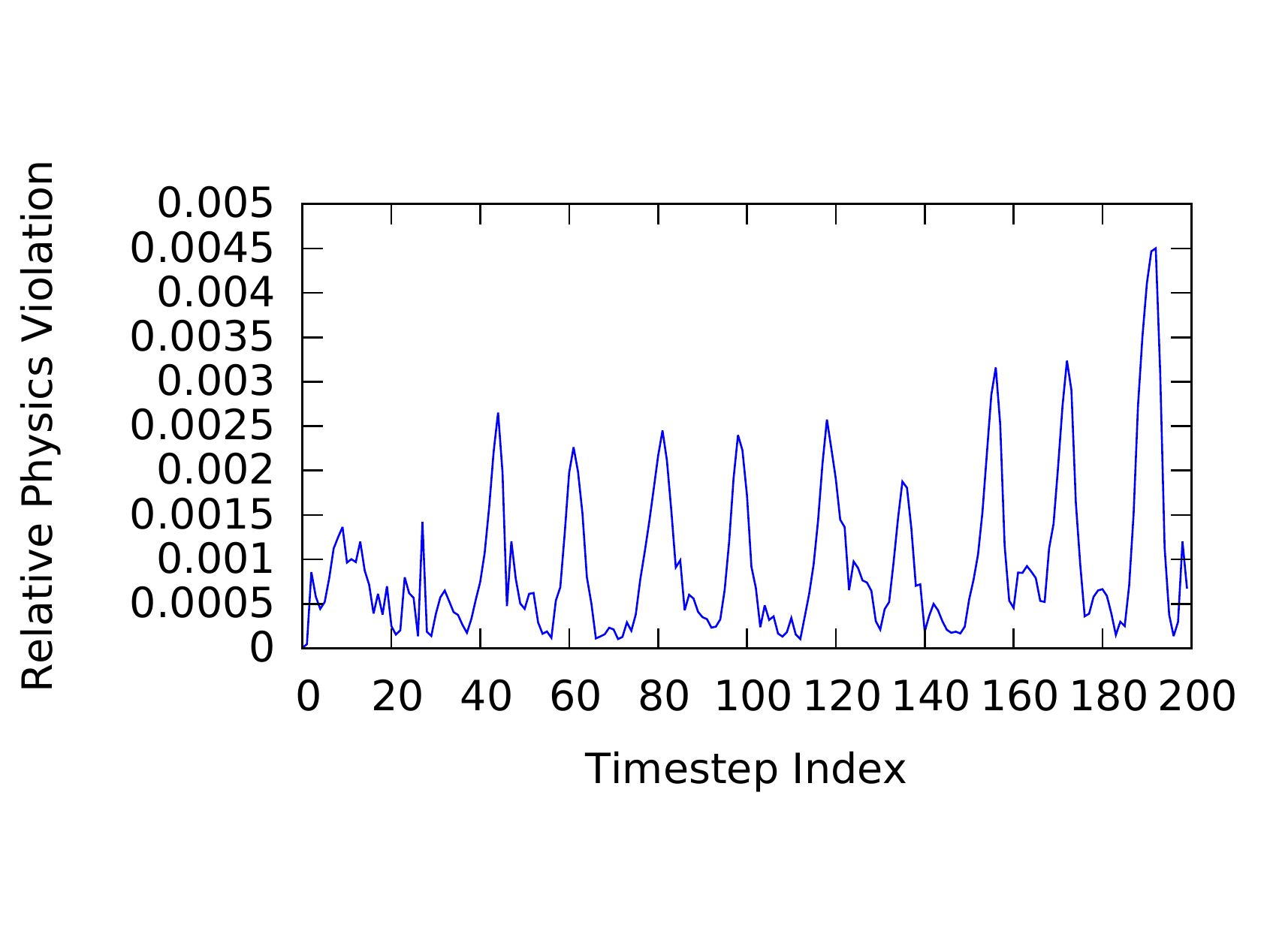} &
\includegraphics[trim={0 2cm 0 2cm},clip,width=0.25\textwidth]{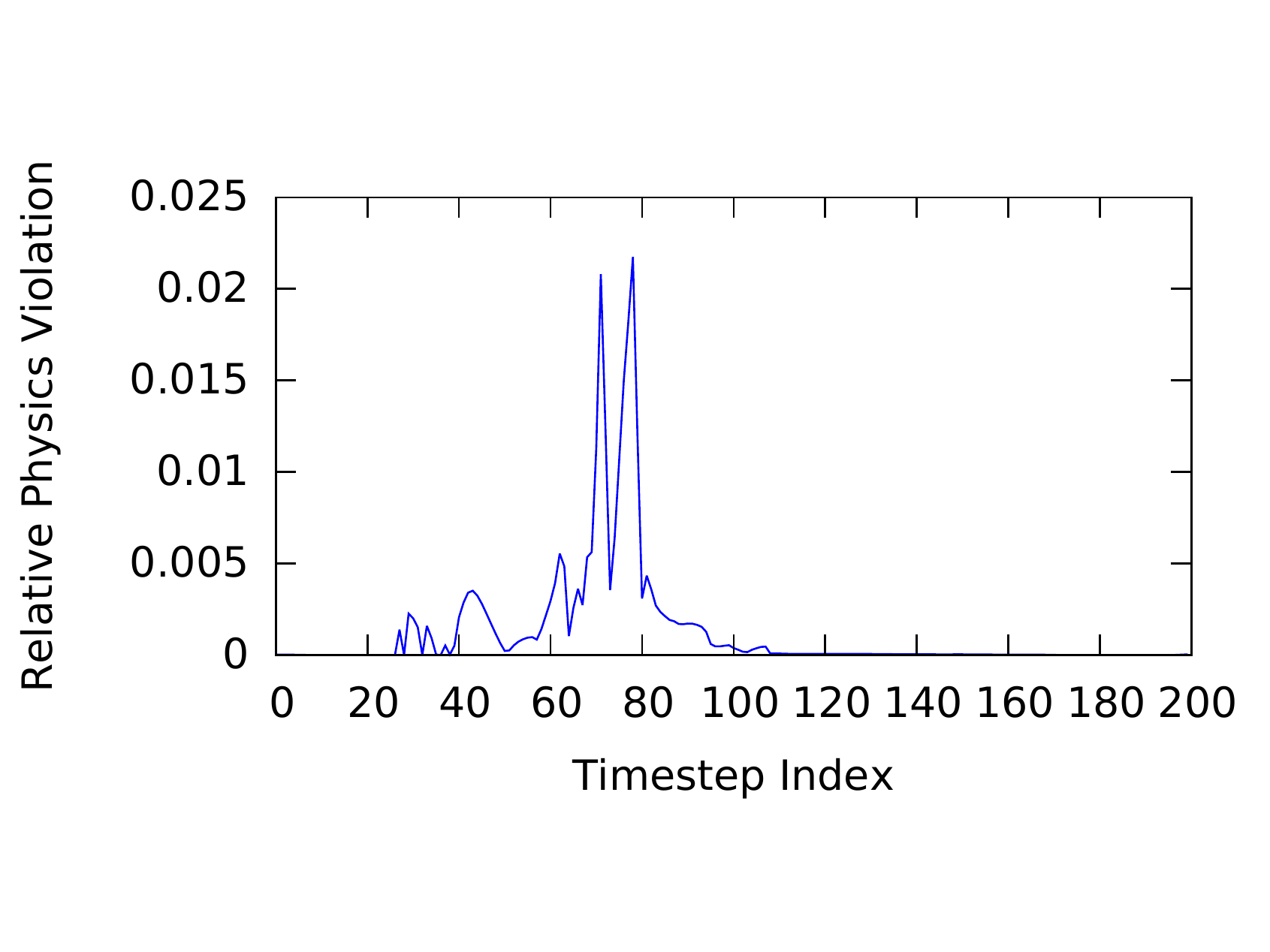} \\
\hline
Cross Rolling \\
\hline
\includegraphics[trim={0 2cm 0 2cm},clip,width=0.25\textwidth]{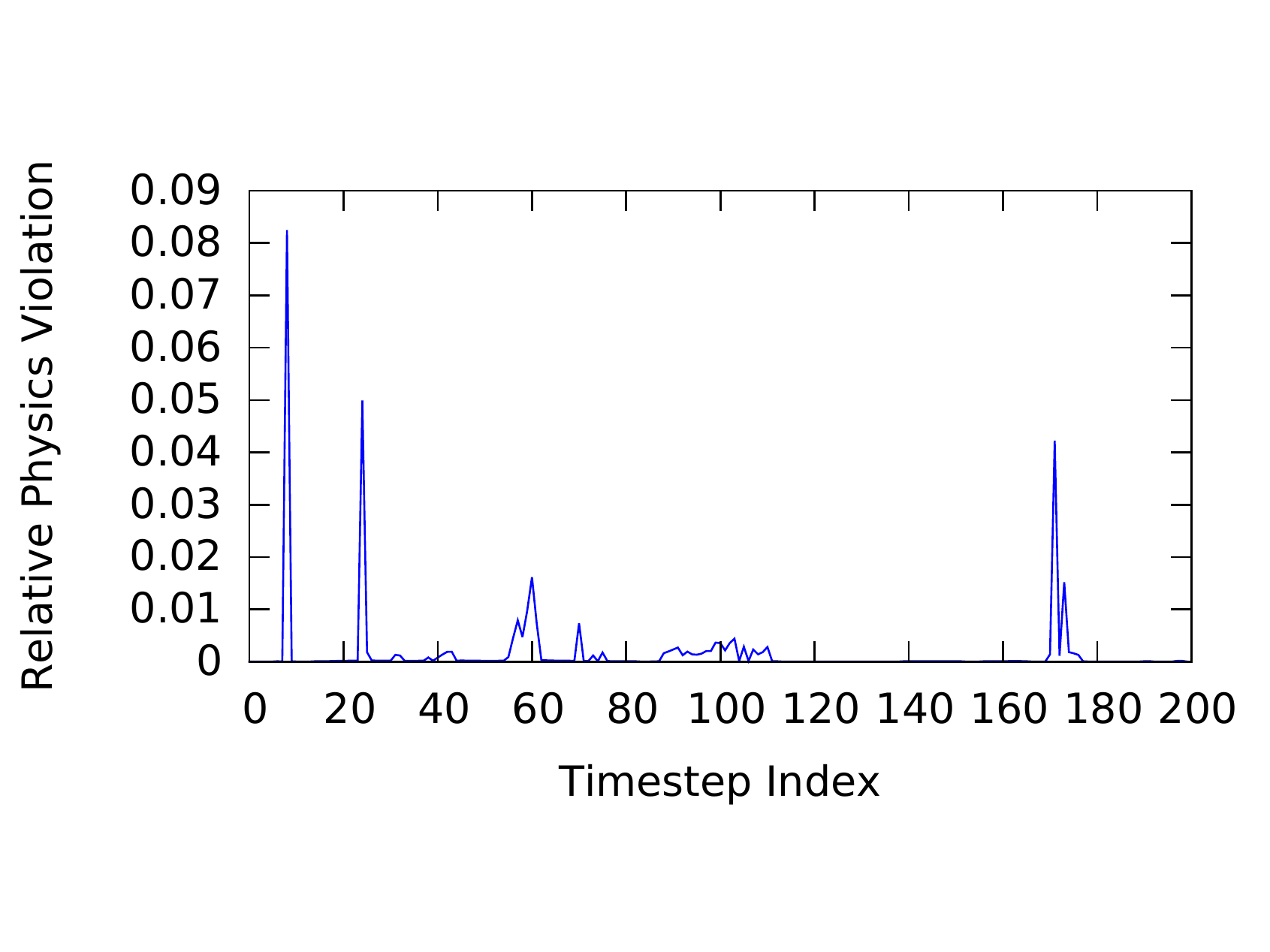} \\
\hline
\end{tabular}}
\end{center}
\caption{\label{table:physVio} The physics violation of all the 7 benchmarks.}
\end{table}

\end{document}